\documentclass[aps,prb,twocolumn,superscriptaddress,floatfix,letter]{revtex4}
\usepackage{graphicx,bm,amsmath,amssymb,color}
\renewcommand{\vec}[1]{\bm{#1}}

\begin{document}

\title{Phase diagram of an anisotropic frustrated ferromagnetic
  spin-$\frac{1}{2}$ chain in a magnetic field: a density matrix
  renormalization group study}

\author{F. Heidrich-Meisner}
\affiliation{Institut f\"ur Theoretische Physik C, RWTH Aachen University, 52056 Aachen, Germany, and
JARA -- Fundamentals of Future Information Technology, Research Centre J\"ulich, 52425 J\"ulich, Germany}
\affiliation{Kavli Institute for Theoretical Physics, University of Santa Barbara, California 93106, USA} 

\author{I. P. McCulloch} 
\affiliation{School of Physical Sciences, The University of Queensland, Brisbane, QLD 4072, Australia }

\author{A. K. Kolezhuk}
\thanks{On leave from: Institute of
Magnetism, National Academy of Sciences and Ministry of Education,
03142 Kiev, Ukraine}
\affiliation{Institut f\"ur Theoretische Physik C, RWTH Aachen University, 52056 Aachen, Germany, and
JARA -- Fundamentals of Future Information Technology, Research Centre J\"ulich, 52425 J\"ulich, Germany}

\date{\today}

\begin{abstract}
We study the phase diagram of a frustrated spin-1/2 ferromagnetic
chain with anisotropic exchange interactions in an external magnetic
field, using the density matrix renormalization group method.  We show that
an easy-axis anisotropy enhances the tendency towards
multimagnon bound states, while an easy-plane anisotropy favors
chirally ordered phases.  In particular, a moderate easy-plane
anisotropy gives rise to a quantum phase transition at intermediate
magnetization. We argue that this transition is related to
the finite-field phase transition experimentally observed in the
spin-1/2 compound LiCuVO$_4$. 
\end{abstract}
\pacs{75.10.Pq,75.40.Cx,75.30.Kz,75.40.Mg}

\maketitle
\section{Introduction}

The interplay of frustration and quantum fluctuations in reduced
dimensions often leads to unconventional magnetic order, such as
chiral or spin-nematic states (see, \emph{e.g.},
Refs.\ 
\onlinecite{Villain78,AndreevGrishchuk84,chubukov91a,nersesyan98,kaburagi99,kolezhuk00,%
hikihara01,lecheminant01,kolezhuk05,laeuchli06,vekua07,mcculloch08,hikihara08,okunishi08,sudan09}).
A particularly simple model yet realizing a fascinating variety of
competing phases is the frustrated ferromagnetic spin-$1/2$ chain in
the presence of an external magnetic
field,\cite{hm06a,vekua07,kecke07,hikihara08,sudan09} described by the
Hamiltonian
\begin{eqnarray}
\label{ham}
&& \mathcal{H}= \sum_{l}\Big\{ J_{1}(\vec{S}_{l}\cdot \vec{S}_{l+1})_{\Delta}
+J_{2}(\vec{S}_{l}\cdot \vec{S}_{l+2})_{\Delta} -h S_{l}^{z}\Big\},\nonumber\\
&& (\vec{S}_{1}\cdot \vec{S}_{2})_{\Delta}\equiv  S_{1}^{x} S_{2}^{x}+  S_{1}^{y}
S_{2}^{y}+\Delta  S_{1}^{z} S_{2}^{z}, 
\end{eqnarray}
where $\vec{S}_{l}$ is a spin-$\frac{1}{2}$ operator acting at site $l$,
$J_1<0$ and $J_2>0$ are the nearest and next-nearest neighbor exchange
constants, $h$ is the external magnetic field, and $\Delta$ is the
 exchange anisotropy.  The system may be alternatively viewed as two
 antiferromagnetic chains coupled by a ferromagnetic zigzag-type coupling whose
 strength is measured by 
the frustration parameter 
\begin{equation}
\beta=J_{1}/J_{2}.
\end{equation}
The isotropic ($\Delta=1$)
version of this model has a rich magnetic phase diagram exhibiting states with different
types of competing unconventional
orders.\cite{hm06a,vekua07,kecke07,hikihara08,sudan09} In particular, the vector
chirality, being the quantum remnant of the classical helical spin order,
competes with multipolar orders which characterize the
pseudo-condensate consisting of multimagnon bound states. A similar effect has
been previously predicted\cite{kolezhuk05} and 
recently confirmed numerically\cite{mcculloch08,okunishi08} for the case of
the \emph{antiferromagnetic} frustrated chain with $J_{1}>0$,
$J_{2}>0$.

The \emph{vector chirality (spin current)}
$\vec{\kappa}_{l} = (\vec{S}_{l} \times \vec{S}_{l+1})  $ can,
even in one dimension, exhibit true long-range order (LRO), \emph{i.e.},
the asymptotic value $\kappa_{0}^{2}=\lim_{|n-n'|\to\infty} C_{\kappa}(n,n')$ of the
chirality correlator
\begin{equation}
\label{chir}
C_{\kappa}(n,n') = \langle \kappa_{n}^{z} \kappa_{n'}^{z} \rangle 
\end{equation}
can be finite. In the presence of an external magnetic
field or of a finite anisotropy $\Delta\not=1$, the rotational $SU(2)$ symmetry
is broken down to $U(1)\times Z_{2}$, and the vector chiral order corresponds to the
spontaneous breaking of the discrete $Z_{2}$ (parity) symmetry. 
At a finite
magnetization, the presence of a nonzero vector chirality automatically leads to
the emergence of  \emph{scalar chirality},\cite{Al-Hassanieh+09} defined as a mixed product of three spins
on a triangular plaquette. It has been shown recently \cite{Bulaevskii+08} that in the underlying
electronic system the presence of a scalar chirality always induces charge
currents, leading to orbital antiferromagnetism.

A common feature of the
\emph{multipolar phases} is that the excitations that correspond to a single spin flip
({\it i.e.}, to a change $\Delta S^{z}=\pm1$ of the $z$-component of the total spin)
are gapped, and therefore,  the in-plane spin correlator $\langle
S^{+}_{n}S^{-}_{n'}\rangle$ decays exponentially with the distance $|n-n'|$. This
distinguishes such phases from the usual spin fluid phases (also called the XY1-type, in the
classification due to Schulz\cite{Schulz86})
where the  $\langle
S^{+}_{n}S^{-}_{n'}\rangle$ correlations decay algebraically.
At the same time, the  excitations  with $\Delta S^{z}=\pm2$ are gapless
in the quadrupolar phase, those with $\Delta S^{z}=\pm 3$ are gapless in the
octupolar  phase, etc. 
The  long-range quadrupolar (nematic) order, characterized by the finite
asymptotic value of
the correlator
\begin{equation} 
\label{nematic}
C_{2}(n,n')=\langle S^{+}_{n}S^{+}_{n+1}S^{-}_{n'}S^{-}_{n'+1}\rangle 
\end{equation}
 at $|n-n'|\to\infty$,
would break the $U(1)$ symmetry, such that those correlations
can only be \emph{quasi}-long-range ({\it i.e.}, exhibiting a power-law decay)
in purely one-dimensional (1D) systems,
yet they may develop into a true LRO in real materials where a finite
three-dimensional interaction is always present. The same applies to the higher
multipolar order parameters such as the octupolar (triatic) order defined by
the correlator of the type $C_{3}(n)=\langle S^{+}_{l}S^{+}_{l+1}S^{+}_{l+2}
S^{-}_{l+n}S^{-}_{l+n+1}S^{-}_{l+n+2}\rangle$, etc. Finally, the spin density correlator
\begin{equation} 
\label{sdw} 
C_{\rm SDW}(n,n')=\langle S^{z}_{n}S^{z}_{n'}\rangle-\langle S^{z}_{n}\rangle
\langle S^{z}_{n'}\rangle
\end{equation}
 has a power-law decay in multipolar phases (as well
as in the other phases mentioned above), and depending on the dominant
correlations,  a multipolar phase can be further characterized as being of the
nematic (triatic, etc.) or  spin-density wave (SDW) type.

\begin{figure}[tb]
\includegraphics[width=0.49\textwidth]{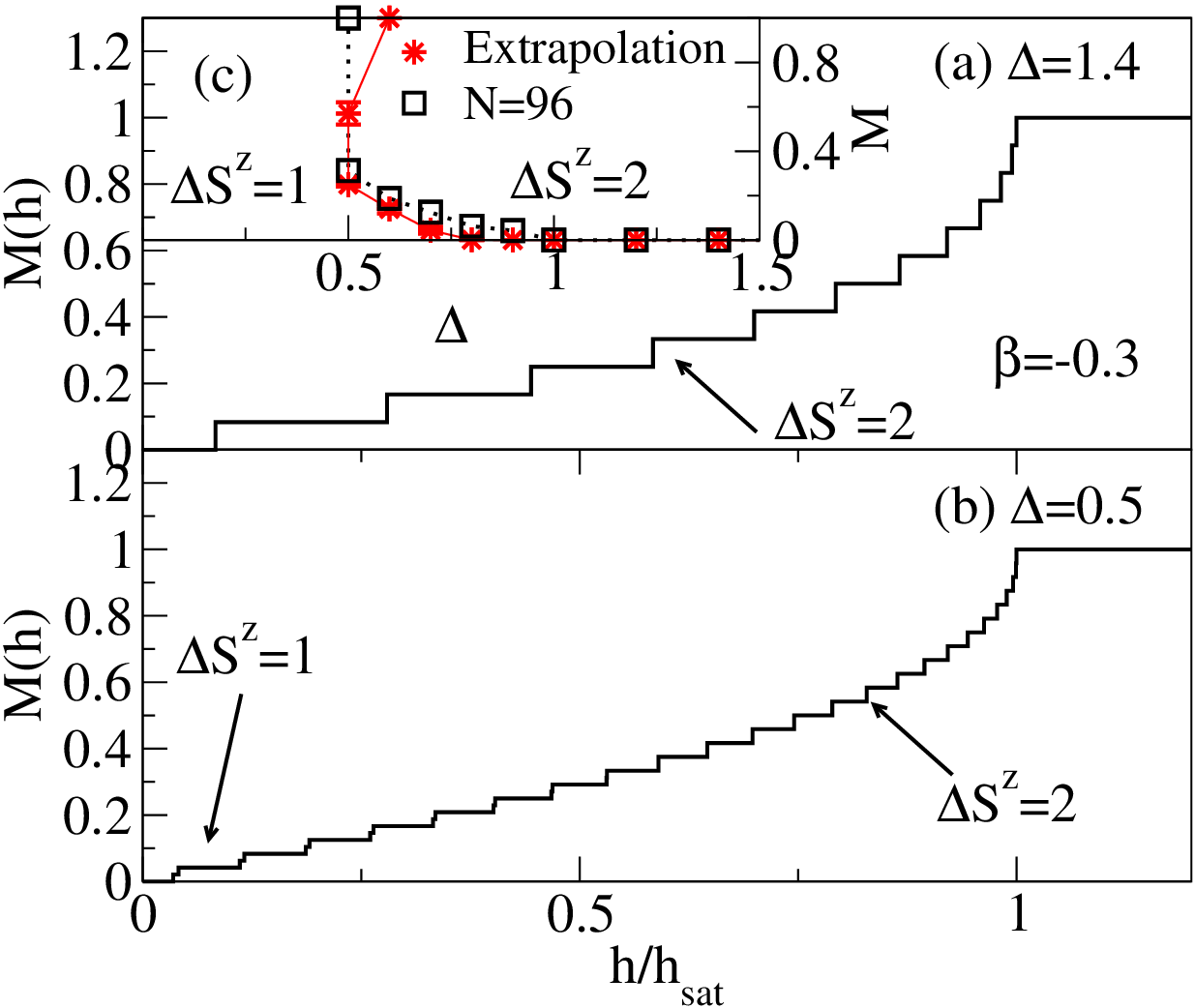}
\caption{(Color online) (a) and (b) Magnetization curves for $\beta=-0.3$ and
  (a) $\Delta=1.4$ and (b) $\Delta=0.5$. (c) Phase boundaries in the $M$ vs
  $\Delta$ plane, based on $M(h)$, for $\beta=-0.3$ (squares: $N=96$; stars:
  extrapolation in $1/N$).  }
\label{fig:phase03}
\end{figure}

\begin{figure}[tb]
\includegraphics[width=0.49\textwidth]{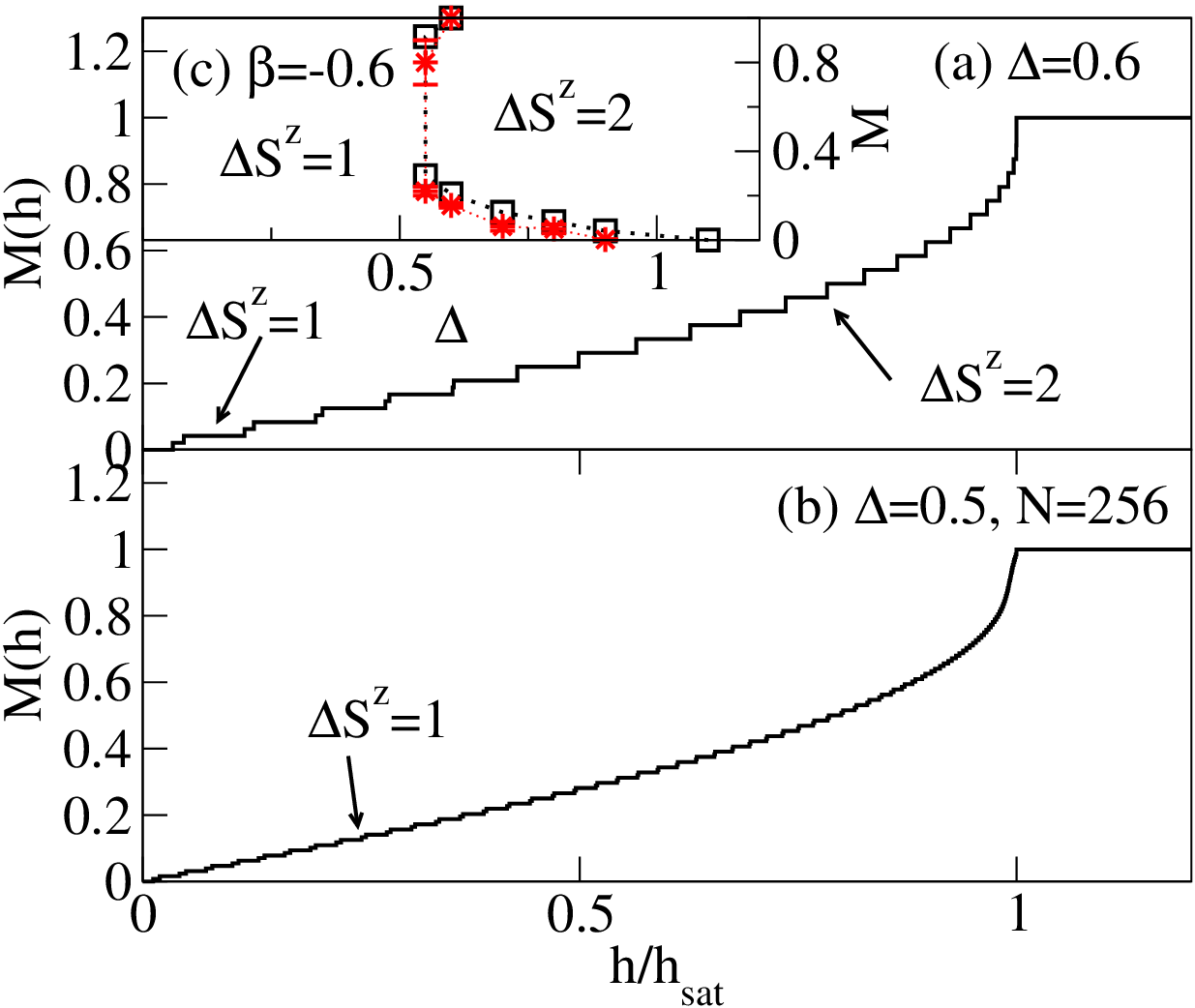}
\caption{(Color online) (a) and (b) Magnetization curves for $\beta=-0.6$ and
  (a) $\Delta=0.6$ and (b) $\Delta=0.5$. (c) Phase boundaries in the $M$ vs
  $\Delta$ plane, based on $M(h)$, for $\beta=-0.6$ (squares: $N=96$; stars:
  extrapolation in $1/N$).  }
\label{fig:phase06}
\end{figure}

\begin{figure}[tb]
\includegraphics[width=0.49\textwidth]{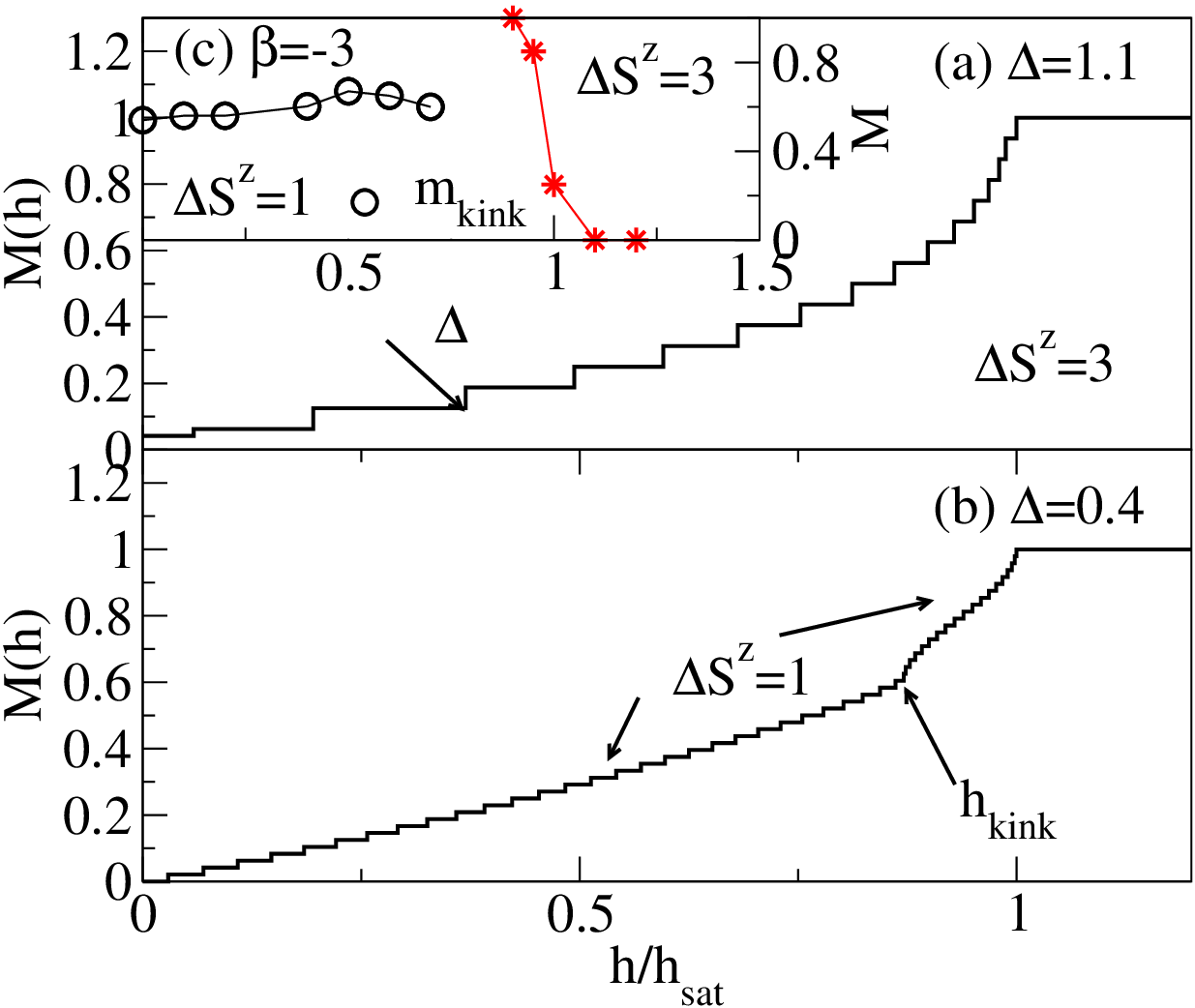}
\caption{(Color online) (a) and (b) Magnetization curves for $\beta=-3$ and (a) $\Delta=1.1$
and (b) $\Delta=0.4$. The arrow in (a) indicates the location where the magnetization
starts to increase in steps of $\Delta S^z =3 $. (c) Phase boundaries in the $M$ vs $\Delta$ plane, based
on $M(h)$,  for $\beta=-3$ (extrapolated in the system size $N\to\infty$).
 }
\label{fig:phase3}
\end{figure}

In the
isotropic model at small $|\beta|$ the spin gap is predicted to be either
zero\cite{WhiteAffleck96,AllenSenechal97} or astronomically small.\cite{ItoiQin01}
The zero-field phase diagram of the frustrated ferromagnetic chain with an
anisotropic exchange has been studied, both for the case of anisotropic
nearest-neighbor interactions only\cite{plekhanov08,avella08} and for the case
in which both exchange paths exhibit the same
anisotropy.\cite{somma01,dmitriev08,dmitriev09,furukawa08} For the latter
example, relevant to our work, the existence of dimer, spin fluid, and
(anti)ferromagnetically ordered phases has been suggested.\cite{somma01}
Moreover, a chirally ordered phase has been predicted to exist at $|\beta|
\lesssim 1$ for $\Delta <1$.\cite{furukawa08}

The model \eqref{ham} has been suggested to be relevant for the description of
several recently discovered quasi one-dimensional magnetic materials such as
LiCuVO$_4$ (Ref.~\onlinecite{enderle05}), Rb$_2$Cu$_2$Mo$_{3}$O$_{12}$
(Ref.~\onlinecite{hase04}), Li$_2$ZrCuO$_4$ (Ref.~\onlinecite{drechsler07}), and
anhydrous $\rm CuCl_{2}$ (Ref.\ \onlinecite{banks+09}).

Our goal is to study the interplay between the exchange anisotropy and
the magnetic field as reflected in the magnetic phase diagram of the
model \eqref{ham}.  The motivation for our work stems from the
experimental results\cite{enderle05,banks07,schrettle08} for
LiCuVO$_4$ that have revealed the existence of a phase transition in a
magnetic field from a helically ordered state at low field values into
another phase at high magnetic fields where the magnetic order seems
to be collinear and directed along the field
axis.\cite{Buettgen+07,schrettle08} If one imagines ``switching off''
the three-dimensional interactions, the helical phase might get
transformed either into the chirally ordered phase or into a usual
spin-fluid XY phase (albeit with incommensurate spin correlations),
while the unknown high-field phase could correspond to the quadrupolar
state of the purely 1D model.

However, for the specific parameter values suggested to be relevant for this
particular material, \emph{i.e.}, $\beta\approx -0.3 $, the 1D model \eqref{ham}
with isotropic interactions ($\Delta=1$) does not support any phase transitions
at intermediate field values.\cite{hm06a,banks07} Numerical
results\protect\cite{hikihara08,sudan09} for $|\beta|>1$ suggest that the vector
chiral phase shrinks very fast with decreasing $|\beta|$, and thus it is hardly
detectable already at $\beta\simeq -1$. Although one might assume that the
vector chiral phase still persists in an infinitesimally narrow region that
vanishes asymptotically at $\beta\to 0$, this would not suffice to explain the
finite-field transition in LiCuVO$_4$ occuring at a relatively high field
strength of about 20\% of the saturation field.\cite{banks07}

At the same time, electron spin resonance
experiments\cite{Vasilev+01,KrugVonNidda+02} indicate that the exchange
interactions in  LiCuVO$_4$  have
an easy-plane anisotropy of about 10\%. This puts forward a natural question whether including this
type of an anisotropy may drive the sought-for phase transition.  We show that
this is indeed the case: there is a finite window of $\Delta<1$ where 
the spin fluid phase persists at low fields, while the quadrupolar-SDW
state occupies the high-field region.

In the present study, we focus on  parameter values relevant for LiCuVO$_4$
(Ref.~\onlinecite{enderle05}), anhydrous $\rm
CuCl_{2}$ (Ref.\ \onlinecite{banks+09}),
  and Rb$_2$Cu$_2$Mo$_{3}$O$_{12}$
(Ref.~\onlinecite{hase04}), namely $\beta=-0.3$, $-0.6$ and $-3$,
respectively. 
To carry out the numerical analysis, we employ the density matrix
renormalization group (DMRG) method,\cite{white92b,white93,schollwoeck05} and our study
is mainly based on the calculation of magnetization curves $M=M(h)$ and the
chiral order parameter $\kappa_{0}$.  
In Sec.~\ref{sec:mag}, we present the analysis of 
 magnetization curves $M=M(h)$ as a
 function of the exchange anisotropy $\Delta$. From the magnetization curves, we
 are able to extract the phase boundaries. The results of the
 magnetization curves analysis are further supported and supplemented
 by the analysis of correlations presented in Sec.\
 \ref{sec:chiral}. 
 Our main result, the magnetic phase diagrams derived from the
 combined analysis of magnetization curves and correlations functions,
 is presented in Sec.~\ref{subsec:diags}. We conclude with a summary
 and discussion in Sec.~\ref{sec:sum}.


\section{Magnetization curves}
\label{sec:mag}

In this Section, we present  magnetization curves of the ferromagnetic
frustrated chain and discuss
their relation to the phase boundaries  of the model (\ref{ham})
in the magnetization  vs
anisotropy plane.  
To that end, we compute the ground-state
energies $E_0(S^z)$ for all values $S^z=\sum_l S^z_l$ of the $z$-component of the total spin.
By subtracting the Zeeman energy $-h S^z$ and
carrying out the Maxwell construction, we find, for each given field $h$,  the
quantum number $S^z$ and respectively the magnetization $M=2S^z/N$
of the ground state, where $N$ is the number of sites. Typically, we use about $m=600$ DMRG states, and
open boundary conditions. 

Our results for $\beta=-0.3$, $-0.6$, and $-3$
are shown in
Figs.~\ref{fig:phase03}, \ref{fig:phase06}, and \ref{fig:phase3},
respectively. For both $\beta=-0.3$ and $\beta=-0.6$, at $\Delta=1$ the
system is in the quadrupolar phase\cite{okunishi03,hm06a,banks07}
(also called ``even-odd'',\cite{okunishi03} or XY2 phase in the
classification by Schulz\cite{Schulz86}): the magnetization
increases in steps of $\Delta S^z=2$, due to the presence of two-magnon bound
states. In this phase, the $\Delta S^z=2$ sector, corresponding to the
simultaneous flip of two spins, is gapless, while single-spin excitations with
$\Delta S^z=1$ are gapped.\cite{vekua07} In terms of correlation functions, at
small fields the leading instability is in the SDW channel, while at high
fields the quadrupolar (nematic) correlations of the type (\ref{nematic})
dominate.\cite{chubukov91a,vekua07}

The results of Figs.~\ref{fig:phase03} and \ref{fig:phase06} show that
an 
easy-axis anisotropy $\Delta>1$ simply
stabilizes the $\Delta S^z=2$ phase [see, {\it e.g.},
  Fig.~\ref{fig:phase03}(a)].\cite{note1}  In contrast to that, an easy-plane anisotropy,
$\Delta<1$, disfavors the formation of two-magnon bound
states and eventually, we observe the disappearance of the $\Delta S^z=2$ region
[see Figs.~\ref{fig:phase03}(b), \ref{fig:phase06}(a) and \ref{fig:phase06}(b)],
giving room to the phase with $\Delta S^z=1$. In Sec.~\ref{sec:chiral}, we will
see that this region exhibits chiral order. The results of the analysis of
$M(h)$ for $|J_1|<J_2$ are summarized in Figs.~\ref{fig:phase03}(c) and
\ref{fig:phase06}(c): in both cases, below $\Delta \lesssim 0.5$, the quadrupolar
phase has disappeared. 
 It is worthwhile to remark that in the case of a weak coupling ($\beta=-0.6$
 and $\beta=-0.3$), we observe a
reentrant behavior in the vicinity of $\Delta \sim 0.55$: as the magnetization
increases, one starts in the $\Delta S^z=1$ region, then enters into the
quadrupolar phase, and reenters into the $\Delta S^z=1$ one at $M\approx 0.75$. As
we shall see below, this picture is also supported by the behavior of the
chirality correlations.

In the vicinity of the saturation field ($M=1$) the position of the boundary of 
the quadrupolar phase  is in  good agreement
with the analysis of Ref.\ \onlinecite{KuzianDrechsler07}.  According to
Ref.\ \onlinecite{KuzianDrechsler07}, the field $h_{s,2m}$ at which the
two-magnon bound state gap closes is given by
$h_{s,2m}/J_{2}=[1+(\Delta+1)^{2}-\Delta^{2}(1-\beta)^{2}]/[2(1-\beta\Delta)]$,
while the respective value for one-magnon states is given by
$h_{s,1m}/J_{2}=(\Delta-1)(1+\beta)+(4+\beta)^{2}/8$.  Comparing those two
fields, one finds, for example, that for $\beta=-0.3$ the instability of the
fully polarized state at the saturation field is by condensation of the
two-magnon bound states at $\Delta>\Delta_{s}\simeq 0.54$, and  by one-magnon
states below that value. The critical point $\Delta_{s}$ is only slightly
dependent on the frustration $\beta$, \emph{e.g.,} at $\beta=-0.6$ one has
$\Delta_{s}\simeq0.58$.

Let us now turn to the regime of strong coupling, $\beta=-3$. In the isotropic case, the system is
in a chiral phase at small magnetizations, and with increasing $M$ one enters a
multipolar (actually, octupolar)
phase.\cite{sudan09} This octupolar phase is characterized by $\Delta S^z=3$ steps in the
magnetization curve,\cite{hm06a} which indicates that
three-magnon bound states are excitations with the lowest energy per
unit of $\Delta S^z$.
Similar to the $|\beta|<1$ case, an easy-axis anisotropy
$\Delta>1$  stabilizes the $\Delta
S^z=3$ multipolar phase. We illustrate this behavior in Fig.~\ref{fig:phase3}(a), showing the
magnetization curve for $\Delta=1.1$.
In the easy-plane region  $\Delta<1$, the magnetization curve further exhibits a
kink-like feature at about $M\sim 0.6$, as the example of $\Delta=0.4$, plotted
in Fig.~\ref{fig:phase3}(b), shows.  We trace this kink back to the
incommensurability and the emergence of multiple Fermi points, following the
reasoning of Refs.~\onlinecite{okunishi99,okunishi03}. The resulting phase diagram
for $\beta=-3$, based on the $M(h)$ curves, is presented in Fig.~\ref{fig:phase3}(c).

Summarizing the results of this section, one can say that the main feature,
common for all values of the frustration parameter $\beta$ considered here, is that
an easy-plane anisotropy $\Delta <1$ gives rise to a mid-field phase transition
from the $\Delta S^z=1$ ``phase'' at low fields to a multipolar ($\Delta
S^{z}\geq 2$) phase at high fields.  In Sec.~\ref{sec:chiral}, we will further
focus on characterizing the region with gapless triplet excitations
(\emph{i.e.,} $\Delta S^z=1$) and show that it actually contains several
different phases.

\begin{figure}[tb]
\includegraphics[width=0.45\textwidth]{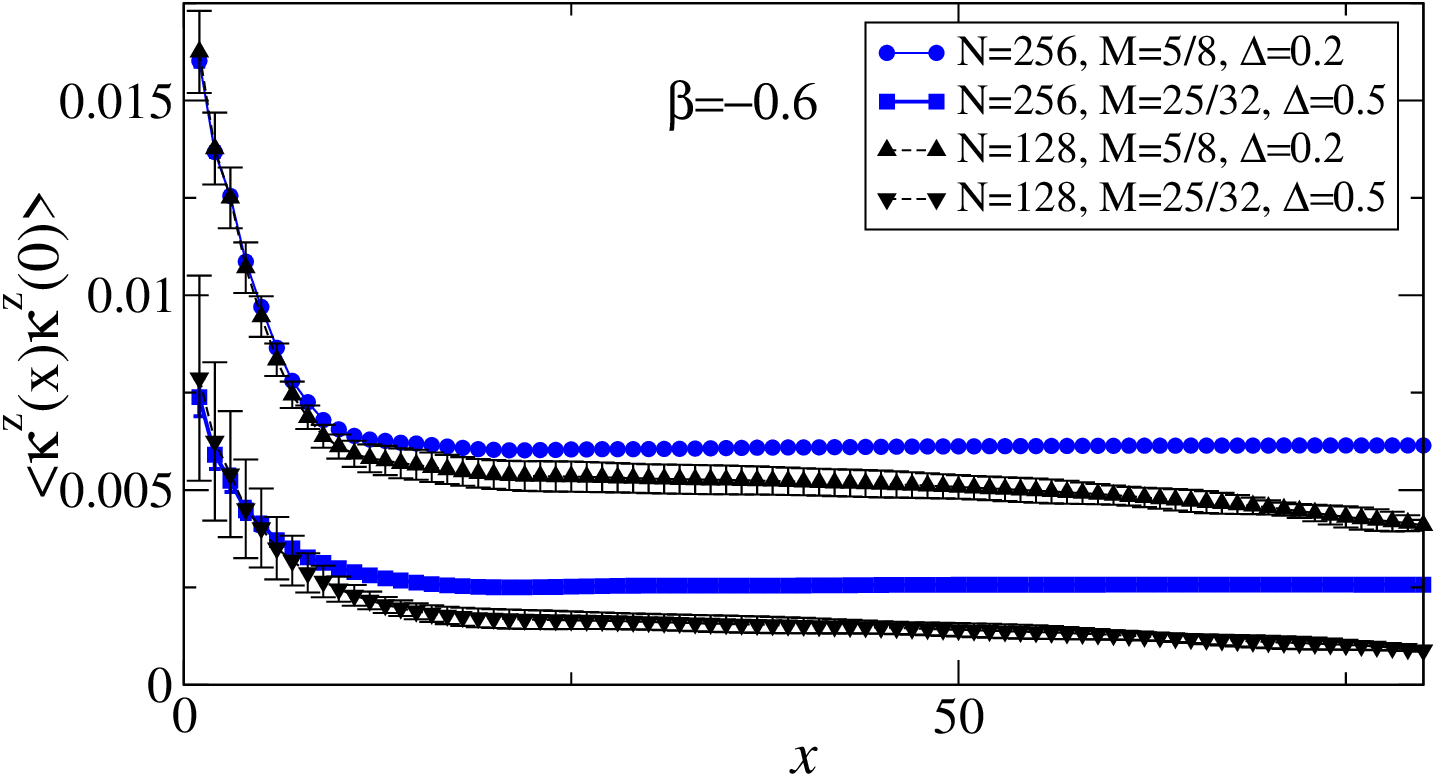}
\caption{(Color online) Examples of the chirality correlation function $\langle \kappa^{z}(x)\kappa^{z}(0)\rangle$ 
at $\beta=-0.6$ for chains with $128$ and $256$ sites, for two fixed points in
the $(\Delta,M)$ plane. }
\label{fig:kz-corr}
\end{figure}


\section{Correlation functions}
\label{sec:chiral}

In this section, we study the  correlation functions,
complementing the analysis of the
magnetization curves presented  in the previous section. 
While the multipolar phases are most easily
detected by the appearance of the $\Delta S^{z} >1$ steps in the
$M(h)$ curves, the region corresponding to $\Delta S^{z}=1$ can
actually contain several different phases. Indeed, 
the  spin-fluid (XY1) phase of the easy-plane spin chain, described
by the one-component
Tomonaga-Luttinger (TL1) liquid, the two-component (TL2) spin-fluid
phase,\cite{okunishi99} and the chirally ordered phase all have gapless
excitations in the $\Delta S^{z}=\pm1$ channel. Thus  all those phases will show up as a single
$\Delta S^{z}=1$ ``phase'' and cannot be further discerned from the $M(h)$ studies.
Analyzing the  chiral correlation
function $C_{\kappa}(n,n')$, we can
identify the chiral phase, and the rest of the $\Delta S^{z}=1$ region
can be divided into the TL1 and TL2 phases by the line where a kink occurs in
the magnetization curve (see Sec.\ \ref{sec:mag}).

Within the multipolar ($\Delta S^{z}=p\geq 2$) phases, an additional analysis of
correlations is necessary to distinguish between the regions with dominant spin
density wave correlations ($\rm SDW_{p}$-''phases'') and those with dominant
multipolar (nematic, triatic, etc.) ones. Although the transitions between,
{\it e.g.}, $\rm SDW_{2}$ and nematic is only a crossover and not a true phase
transition in our purely one-dimensional model, such an analysis can be  helpful
in understanding what could be the resulting order in a real material, where 
frustrated chains are coupled by a weak three-dimensional interaction.

\begin{figure*}[tb]
\includegraphics[width=0.45\textwidth]{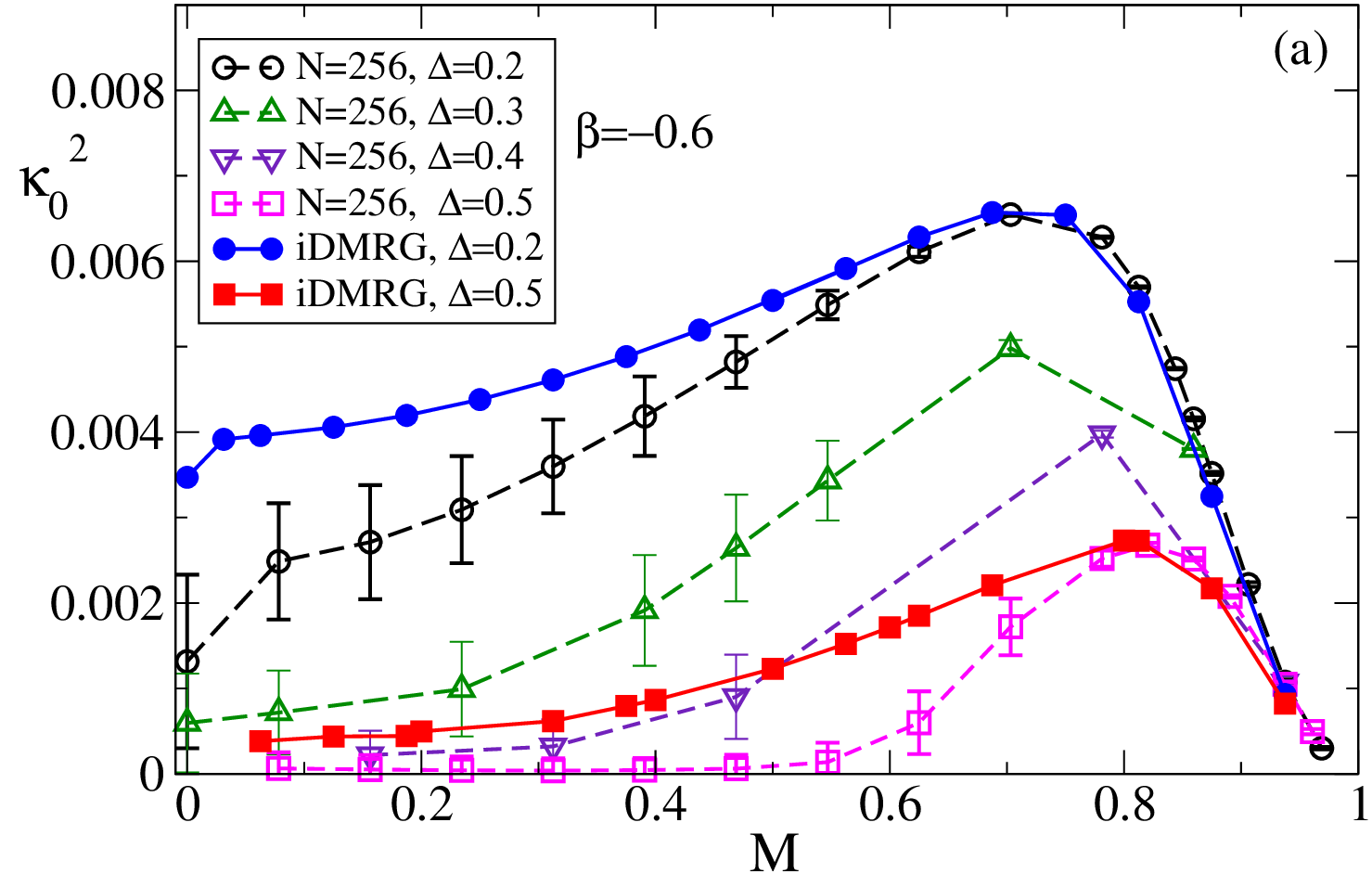}
\includegraphics[width=0.45\textwidth]{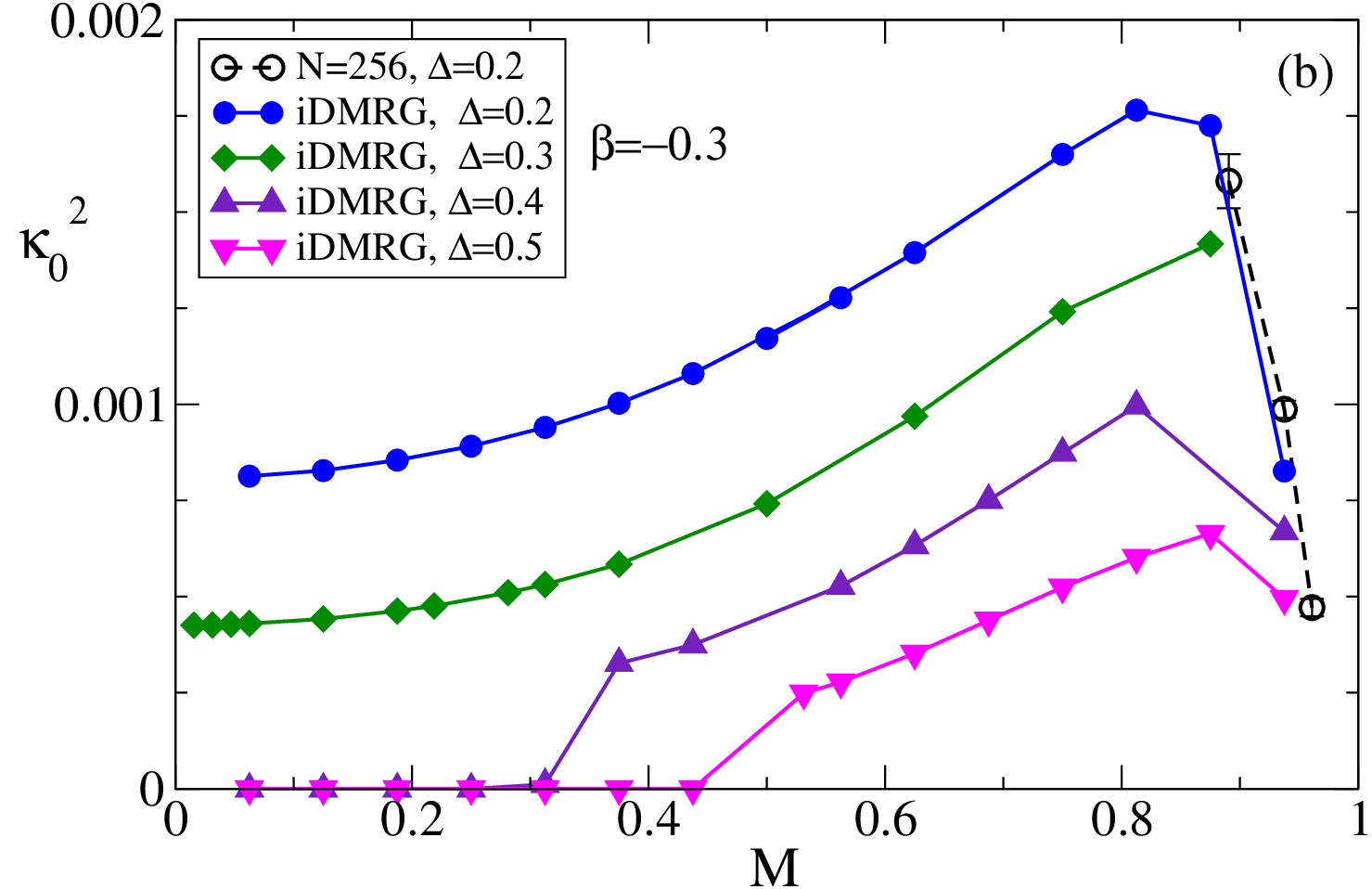}
\vspace*{2mm}
\includegraphics[width=0.45\textwidth]{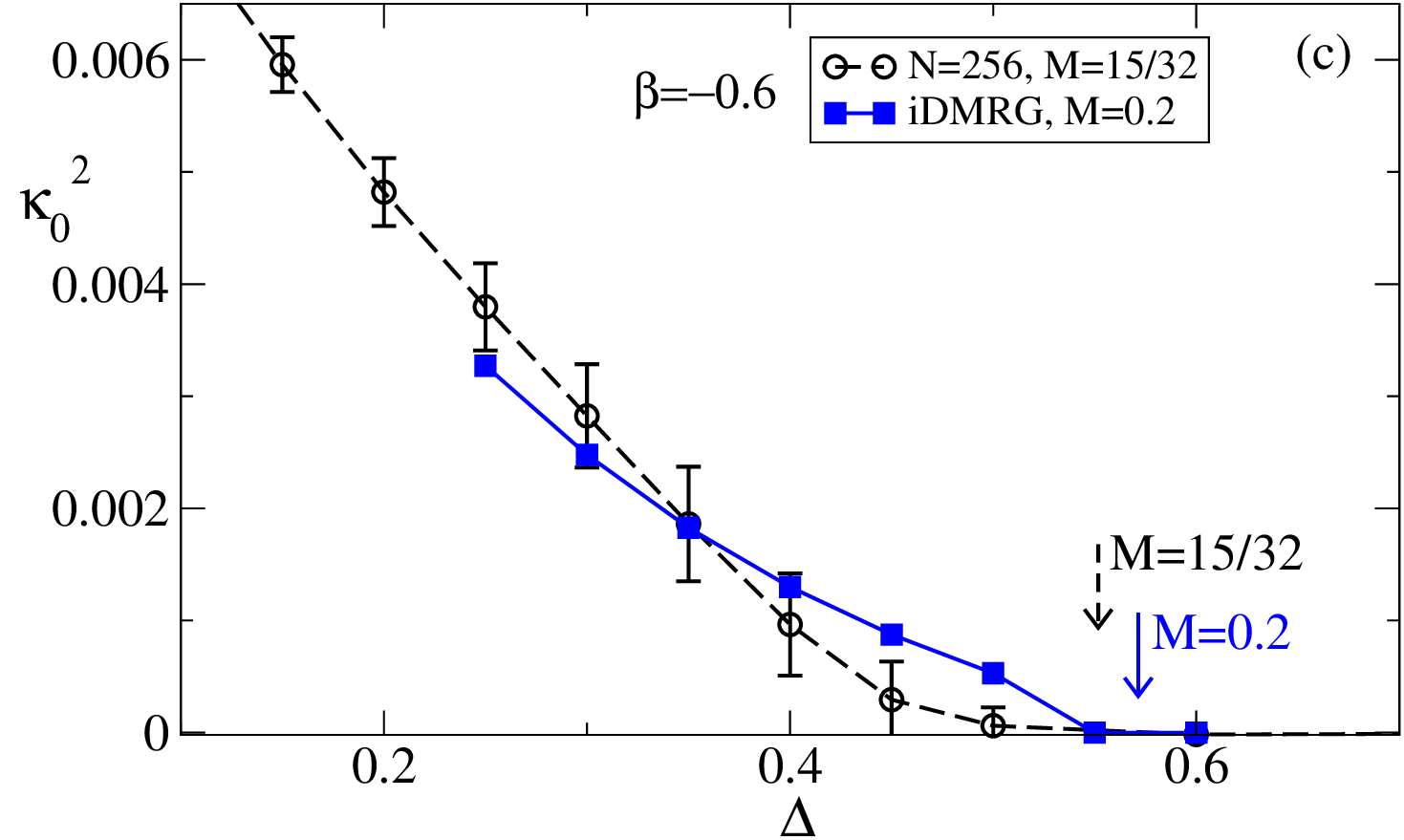}
\includegraphics[width=0.45\textwidth]{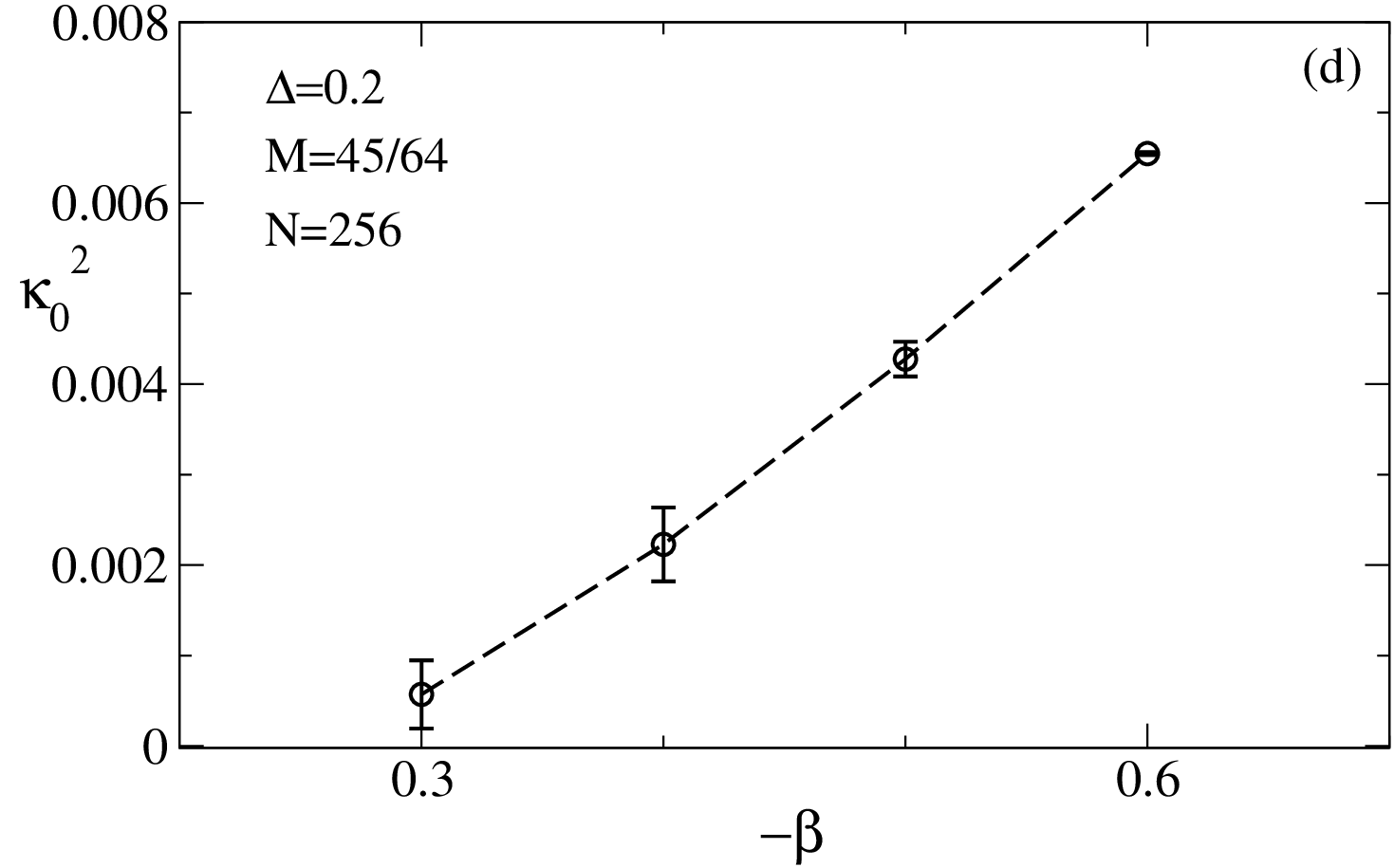}

\caption{(Color online) Square of the vector chirality $\kappa_{0}^{2}$: (a) vs
  magnetization $M$  at $\beta=-0.6$ and various fixed values of the anisotropy $\Delta$; (b) the same at
  $\beta=-0.3$; (c) vs the anisotropy $\Delta$ at two fixed values of $M$ and
  $\beta=-0.6$ (the arrows here denote the respective positions of the $\Delta
  S^{z}=1$ phase boundary);
  (d) vs the  frustration parameter $|\beta|=-J_{1}/J_{2}$ at fixed
  $\Delta=0.2$ and $M=0.703125$. Open symbols denote the finite-size DMRG
  results, and solid symbols correspond to iDMRG.
 }
\label{fig:chirality1}
\end{figure*}

\subsection{DMRG methods for the calculation of the vector chirality}

Using the finite-size DMRG method\cite{white93,schollwoeck05} in its
matrix-product formulation,\cite{McCulloch07} we have studied
correlators (\ref{chir}), (\ref{nematic}), and (\ref{sdw}) in chains
of $N=256$ spins. This length has been chosen since, on the one hand,
it is sufficiently large to study the asymptotic long-distance
behavior of the correlations, and on the other hand, it is small
enough to ensure that the DMRG calculation converges with a moderate
number $m$ of representative states kept.  The typical value of $m$
necessary to reach good convergence strongly depends on the
frustration parameter $\beta=J_{1}/J_{2}$: while for $\beta=-3$,
$m=400$ is normally sufficient, at smaller coupling ($\beta=-0.6$)
this figure grows to $m\simeq 600$-$800$, and in the regime of weakly
coupled chains ($\beta=-0.3$) one needs $m\simeq 800$-$1200$, even for
large magnetizations $M\gtrsim 0.7$ where the convergence is generally
faster.  The correlators (\ref{chir})-(\ref{sdw}) have been calculated
for a large number of ground states in sectors with different
$S^{z}$. They have been averaged over the starting and final positions
$n$, $n'$, and contributions with $n$ or $n'$ being closer as as a
fixed ``cutoff'' (taken here to be $20$ sites) to the chain ends have
been discarded.  Typical chiral correlation functions are shown in
Fig.\ \ref{fig:kz-corr}.  From such data we have extracted the
asymptotic value of the correlator which corresponds to the square of
the chirality $\kappa_{0}^{2}$.

A proper finite-size scaling analysis of chirality correlations is,
however, hampered by strong boundary
effects\cite{mcculloch08,hikihara08,okunishi08} that tend to spoil the
bulk correlations for smaller system sizes. Due to that, it becomes
difficult to distinguish the chiral LRO from a non-chiral phase in
those situations where the chiral order parameter $\kappa_{0}$ becomes
very small.
In such cases, we have complemented the finite-size DMRG study with another
technique, namely, the recently proposed\cite{iDMRG} matrix-product formulation
of the infinite-size DMRG algorithm (iDMRG) which allows to treat systems with
finite magnetization (in contrast to the conventional infinite-size
DMRG method, see, e.g., Ref.\ \onlinecite{hikihara01}). We utilize an algorithm with conserved $U(1)$ symmetry to
constrain the \emph{average} magnetization per unit cell. The convergence rate
of iDMRG is essentially independent of the size of the unit cell, which can be
arbitrarily large.  The advantage of the iDMRG is that the scaling in $(m,N)$ is
replaced by the scaling in the number of states $m$ alone, which can be
translated into a scaling with respect to the correlation length via $\xi \sim
m^\eta$, where the correlation length $\xi$ is determined from the next-leading
eigenvalue of the transfer operator.\cite{criticalscaling}  For critical states
described by a conformal field theory (CFT), $\eta$ is a function of the central charge.\cite{Pollmann}
The spectrum of the transfer operator also gives detailed information about the
exponents and operator content of the CFT.\cite{IanInProgress}

In the standard finite-size DMRG formulation, the degeneracy of two chirally ordered ground states
will be lifted by finite-size corrections. Therefore the purely real ground state of a
finite system is obtained as a superposition of states with
$\pm\kappa_{0}$. The iDMRG,\cite{iDMRG} in contrast,
allows for a spontaneous breaking of the parity symmetry, which also breaks time
reversal symmetry and leads to a complex valued wavefunction. This gives a
transfer operator that is not Hermitian, but is instead complex-symmetric.
The chirality order parameter can then be calculated just as
$\kappa_{0}=\Im\langle S^{+}_{n}S^{-}_{n+1}\rangle$. The iDMRG randomly selects
one of the two ground states, with $\kappa_{0}$ either positive or negative.
For broken symmetry states the iDMRG is quite efficient, because the broken symmetry state
requires fewer basis states than a superposition. 
For example, the representation of a superposition of the $\pm\kappa_{0}$ states
in a form of a matrix product state requires precisely double the number of basis states, because the
reduced density matrices of the two degenerate ground states have no overlap in the thermodynamic limit.
In a finite-size calculation, the mixing of the two states leads to
somewhat less than a factor 2 in the required basis size, nevertheless one still requires generally
fewer states in iDMRG compared with its finite-size counterpart.


\subsection{Vector chirality in the weak coupling regime ($|J_{1}|<J_{2}$)}
\label{sec:chiral_a}

The results for the weak coupling regime $|\beta|<1$ are shown in
Fig.\ \ref{fig:chirality1}. One can see that the finite-size DMRG results give the
 impression that both at $\beta=-0.6$ and $\beta=-0.3$, the vector chiral LRO
vanishes in the low-field part of the $\Delta S^{z}=1$ region. However, as
mentioned above, we cannot reliably detect the presence of a very small chiral
order with the finite-size DMRG method because of strong boundary
effects. 
Applying the iDMRG technique, one can clearly see that the finite-size
DMRG tends to underestimate the value of the chiral order parameter
$\kappa_{0}$, cf. Fig.\ \ref{fig:chirality1}(a,b).\cite{note-errbars}

\begin{figure}[tb]
\includegraphics[width=0.45\textwidth]{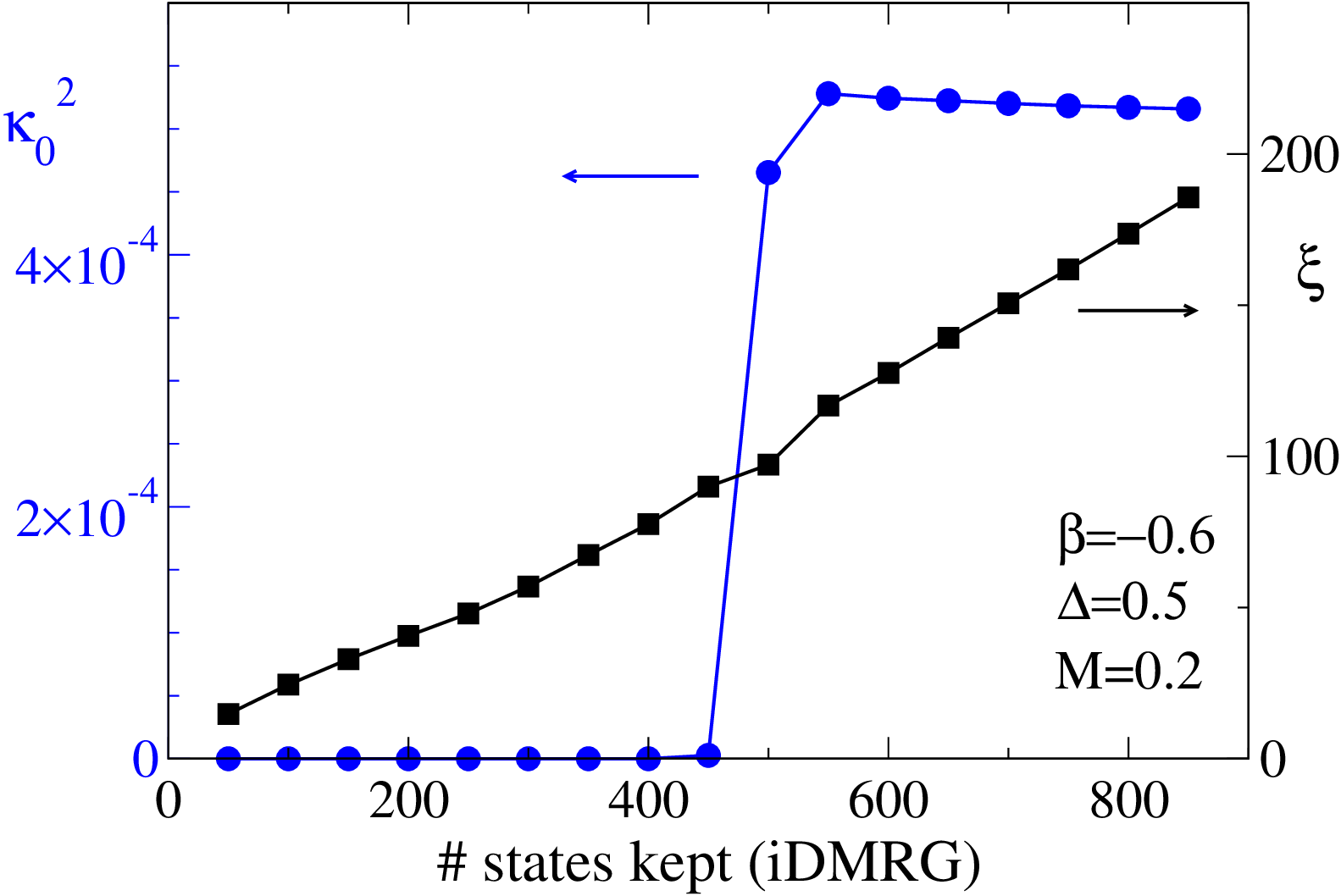}
\caption{(Color online)  Convergence of the iDMRG method at the point
  $\beta=-0.6$, $\Delta=0.5$, $M=0.2$. The largest intrinsic correlation length
  $\xi$ of the matrix-product iDMRG method\protect\cite{iDMRG}  and square of the chirality 
$\kappa_{0}^{2}$ are shown as functions of the number of states kept in
the iDMRG calculation.}
\label{fig:xikappa}
\end{figure}

\begin{figure}[tb]
\includegraphics[width=0.45\textwidth]{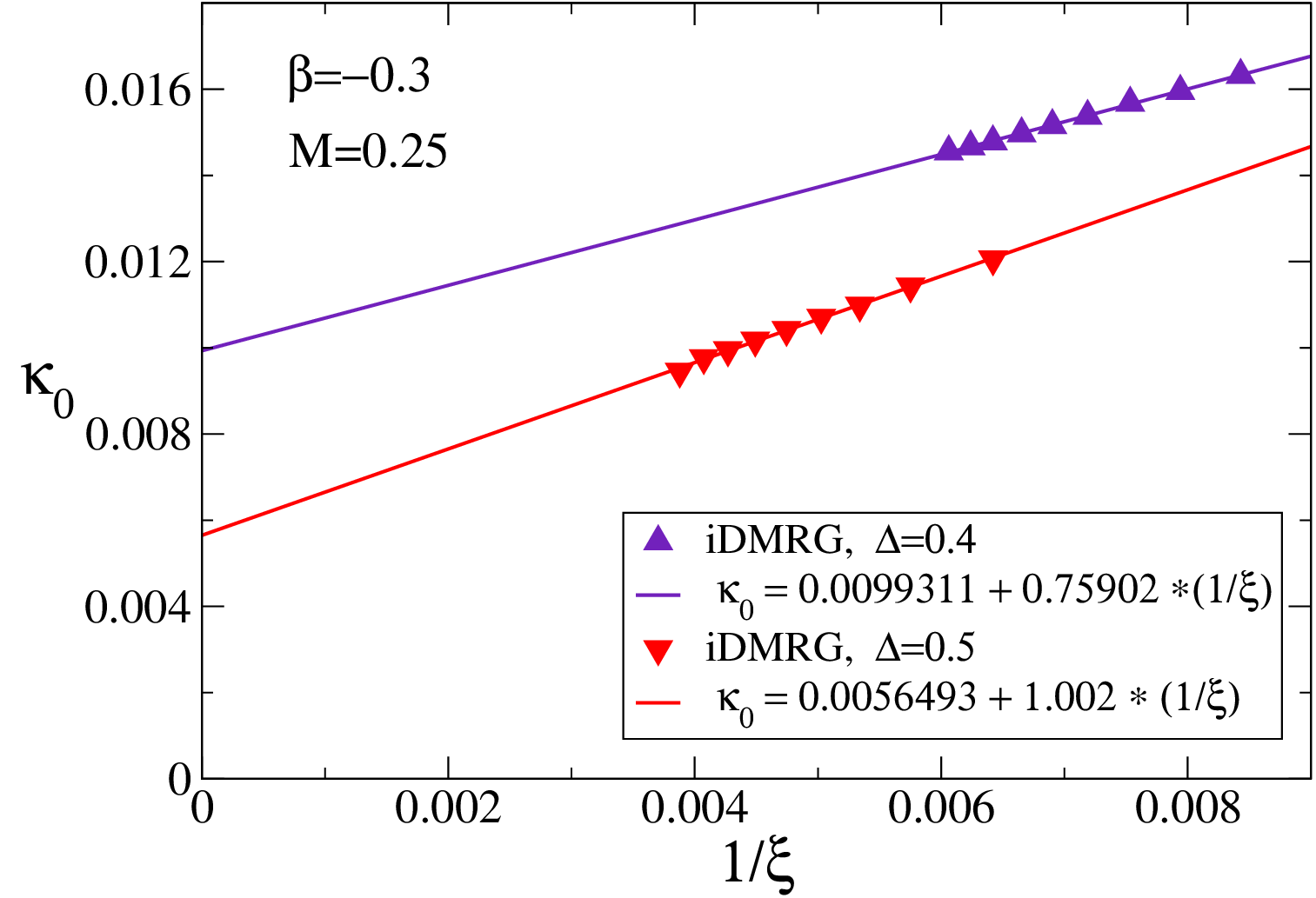}
\caption{(Color online)  An example of extrapolation in the inverse
  correlation length $\xi$ for the chiral order parameter $\kappa_{0}$, at
  $\beta=-0.3$ and $M=0.25$, for two anisotropies $\Delta=0.4$ and $\Delta=0.5$
  where $\kappa_{0}$ becomes very small,
  cf. Fig.\ \protect\ref{fig:chirality1}(b). Up to $m=1000$ steps were kept in
  those iDMRG calculations.
}
\label{fig:kappa-1/xi}
\end{figure}

As can be seen from
Figs.\ \ref{fig:chirality1}(c) and (d), the magnitude of the chiral order parameter
diminishes quickly when $|\beta|$ decreases, and also when one approaches the
boundary of the $\Delta S^{z}=1$ region. The convergence in those cases becomes
very slow. 
Figure \ref{fig:xikappa}
shows the convergence of the iDMRG method at a point close to the $\Delta
S^{z}=1$ boundary: a finite chirality is detected when the largest intrinsic
correlation length $\xi$ of the method exceeds 100 sites. 
Taking guidance from the bosonization picture,\cite{kolezhuk05,hikihara08}
it is fair to assume that the chirality can be detected only after $1/\xi$ drops
below the value corresponding to the spectral gap in the antisymmetric
sector.\cite{note4}
This  gap becomes very small when one approaches the $\Delta
S^{z}=1$ phase boundary, or the $M=0$ line.\cite{ItoiQin01}
In such cases  [see,
\emph{e.g.,} the low-field
region at $\Delta\gtrsim 0.4$ for $\beta=-0.3$ in
Fig.\ \ref{fig:chirality1}(b)], 
one can use an extrapolation  in
$1/\xi$ to extract the  chirality $\kappa_{0}$;
Fig.\ \ref{fig:kappa-1/xi} illustrates that this procedure yields a
finite value of $\kappa_{0}$.
Continuity arguments suggest that the entire $\Delta S^{z}$ region
belongs to the chiral phase, both for $\beta=-0.6$ and $\beta=-0.3$.
This is also consistent with the theoretical prediction\cite{furukawa08}
of a chiral phase  emerging at zero field in a wide range of
$\Delta$ in the limit $|\beta|\to 0$,
based on the analysis of small systems (if the system is in the chiral phase already at
$h=0$,  it is  reasonable to assume that the chirality persists at finite field
as well).

\begin{figure}[tb]
\includegraphics[width=0.45\textwidth]{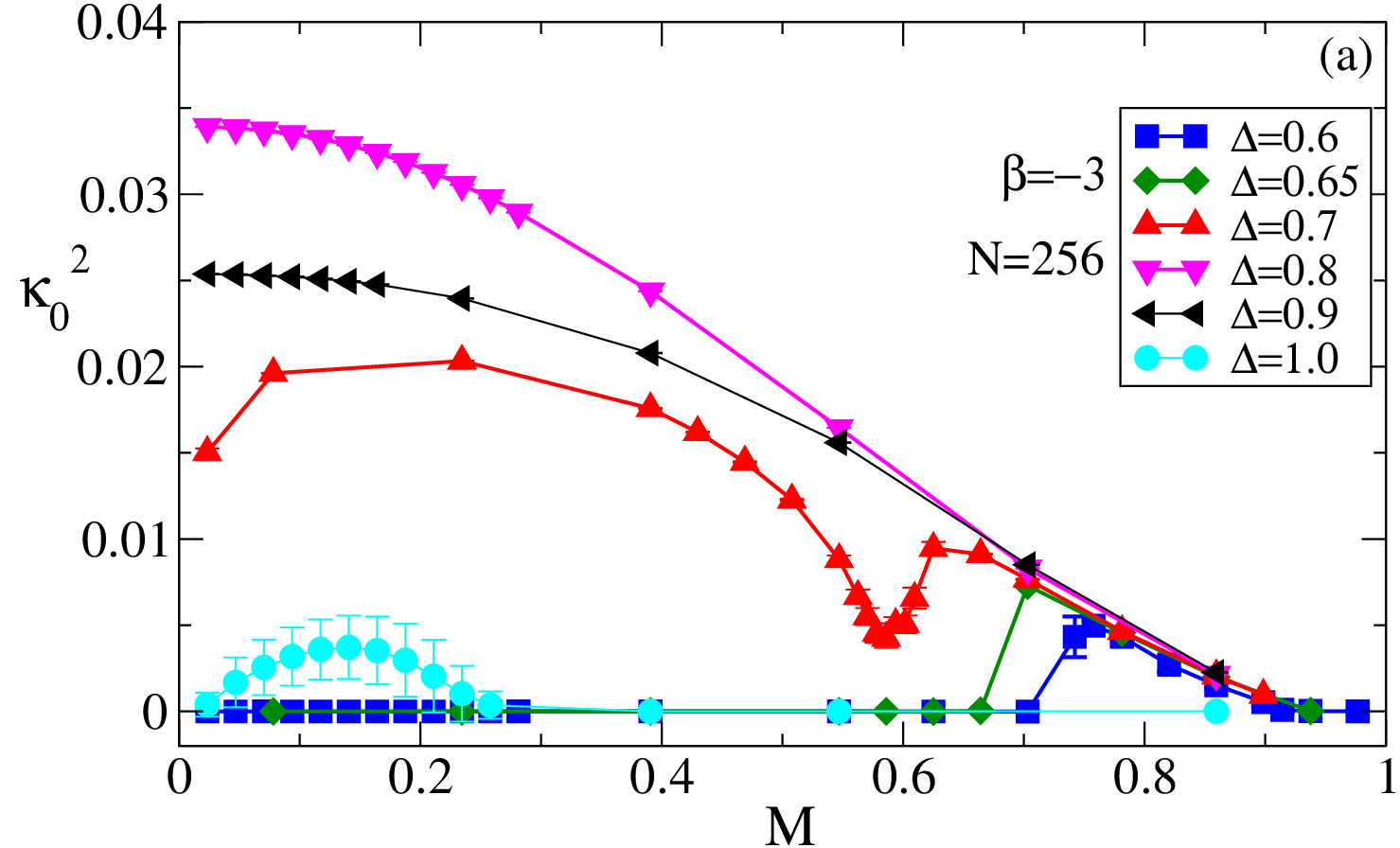}
\vspace*{1mm}
\includegraphics[width=0.45\textwidth]{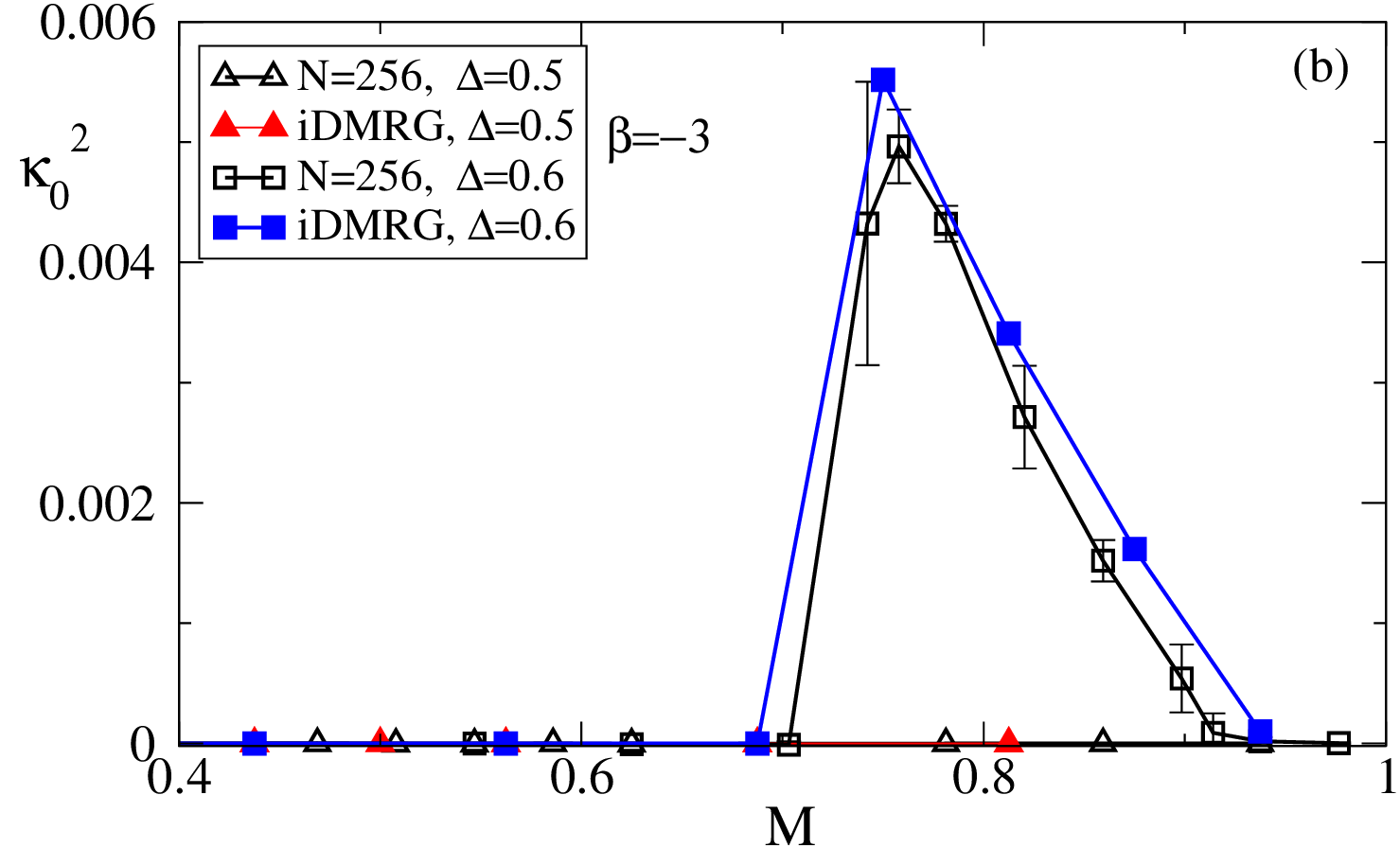}
\caption{(Color online) Square of the vector chirality $\kappa_{0}^{2}$ at $\beta=-3$
as a function of the  magnetization $M$  at several values of the
anisotropy $\Delta$: (a) finite-size DMRG results for a system of
$N=256$ spins; (b) comparison of the finite-size DMRG and iDMRG
results for  paths crossing the chiral phase boundary. }
\label{fig:chirality3}
\end{figure}


\subsection{Vector chirality in the strong coupling regime ($|J_{1}|>J_{2}$)}
\label{sec:chiral_b}

The behavior of the chiral order parameter in the regime of strong coupling
$\beta=-3$, as extracted from the finite-size DMRG and iDMRG calculations, is
shown in Fig.\ \ref{fig:chirality3}.  It indicates the existence of a chiral
phase that is contained inside a relatively narrow stripe $0.5<\Delta<1$, and
the rest of the $\Delta S^{z}=1$ region should belong to a non-chiral
spin-fluid phase.  The presence of a kink in the magnetization curves further
suggests that this spin-fluid phase is in turn divided into the one-component
(TL1) and two-component (TL2) spin fluid phases, occupying the low- and
high-field regions, respectively.

As can be seen from Fig.\ \ref{fig:chirality3}(a), the transition
between the  chiral phase and the TL1 phase is very sharp: the chiral
order drops from a sizeable value to zero in the entire low-field
region, when the anisotropy changes from $\Delta=0.7$ to
$\Delta=0.65$. This suggests that the transition is of the first
order.

\begin{figure}[tb]
\includegraphics[width=0.45\textwidth]{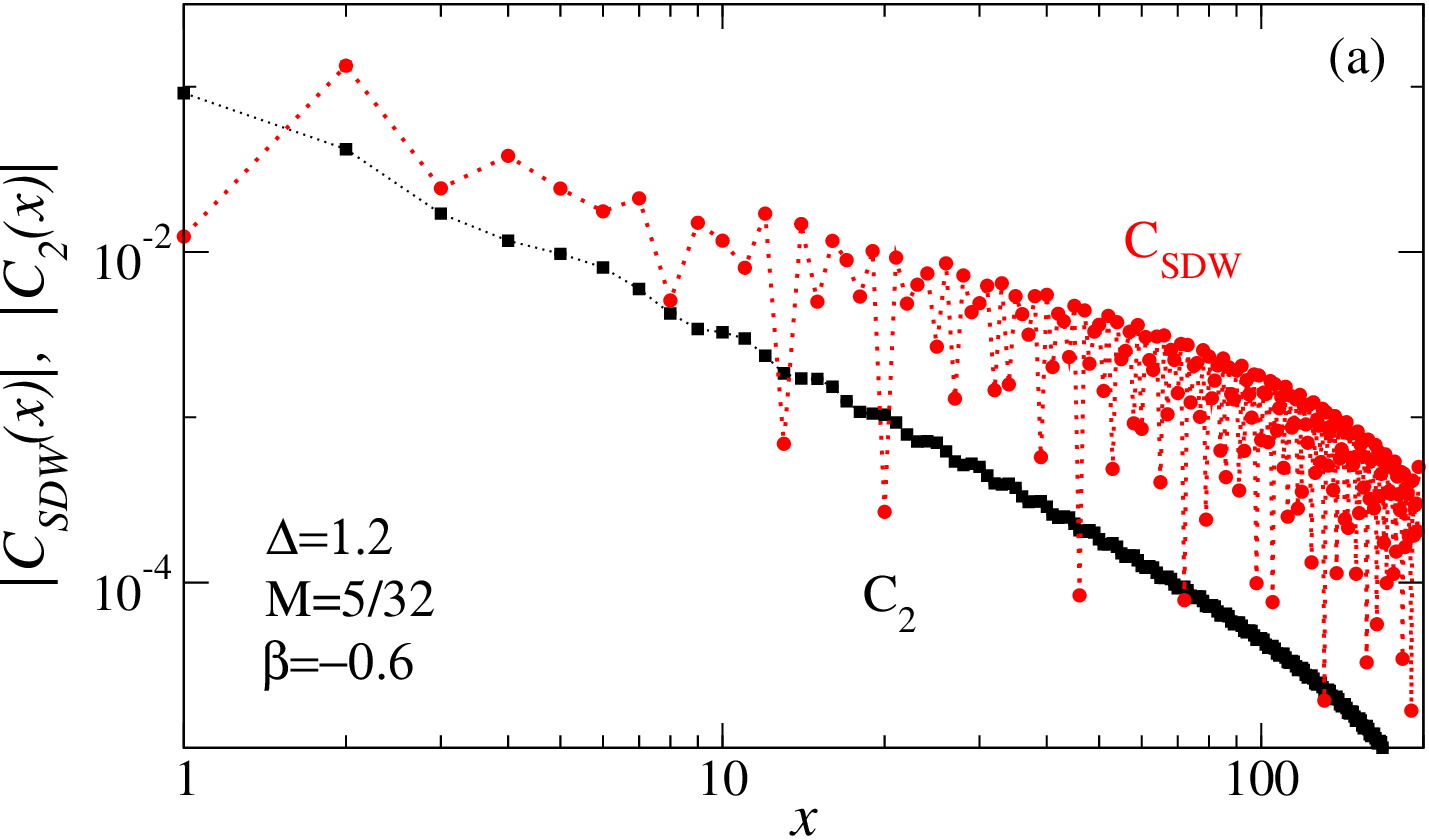}
\vspace*{2mm}
\includegraphics[width=0.45\textwidth]{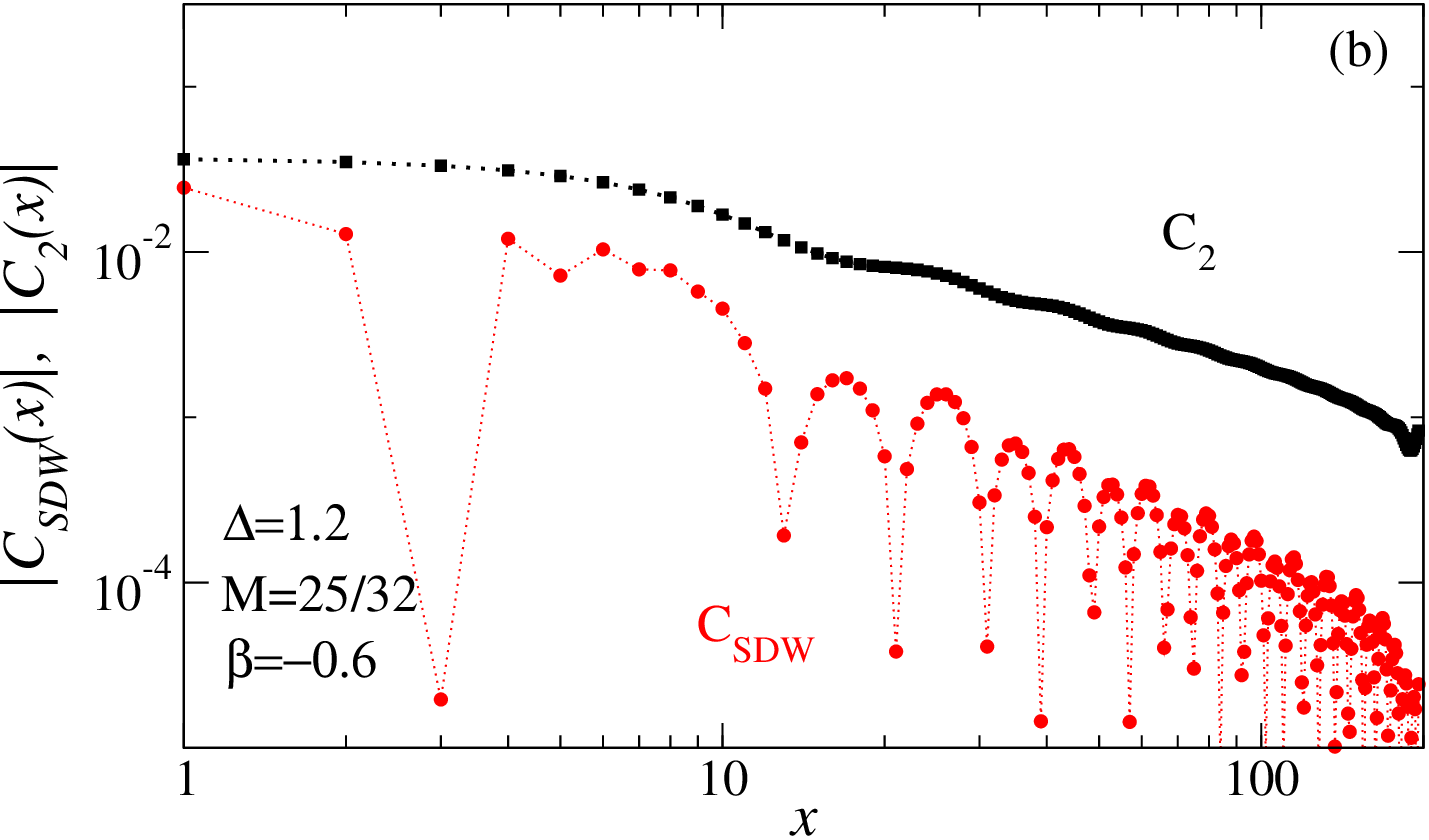}
\caption{(Color online) Comparison of the nematic correlator $C_{2}(x)$ and the
  spin density wave correlator $C_{\rm SDW}(x)$, defined by
  Eqs.\ (\ref{nematic}), (\ref{sdw}), respectively, as a function of distance
  $x=|n-n'|$, at $\beta=-0.6$ and two different magnetizations: (a) $M=5/32$;
  (b) $M=25/32$. Here, finite-size DMRG results for a system
  of $N=256$ spins are shown.}
\label{fig:sdw}
\end{figure}

Figure \ref{fig:chirality3}(b) illustrates that for $\beta=-3$, the
iDMRG results agree very well with the finite-size DMRG data for a $256$-spin
chain. This fact, together with the abrupt character of the transition from the
chiral phase to the TL1 spin fluid, gives us reasons to conclude that in the
strong coupling case the observation of non-chiral regions is not an
artefact of the DMRG convergence, but is due to existence of spin-fluid phases,
in contrast to the behavior in the weak coupling regime $|\beta| <1$.

\subsection{Crossover between the spin density wave and nematic at $\beta=-0.6$}
\label{subsec:sdw}

We have analyzed the crossover between $\rm SDW_{2}$ and  nematic
correlations inside the quadrupolar $\Delta S^{z}=2$ phase at
$\beta=-0.6$. The typical behavior of the SDW and nematic correlations as defined by
Eqs.\ (\ref{nematic}), (\ref{sdw}) is shown in Fig.\ \ref{fig:sdw}. One can see
that both correlators decay as power law, but the SDW correlations dominate in
the low-field
region, while the nematic correlations take over at high magnetizations, in
agreement with the bosonization analysis and earlier numerical
results.\cite{hm06a,vekua07,kecke07}
The effect of the anisotropy $\Delta$ on this crossover is rather mild: an
easy-plane anisotropy $\Delta<1$ shifts the
crossover boundary towards higher $M$,  making the nematic region
more narrow, and the crossover boundary seems to be  insensitive to an
easy-axis anisotropy $\Delta>1$.

\begin{figure}
\includegraphics[width=0.45\textwidth]{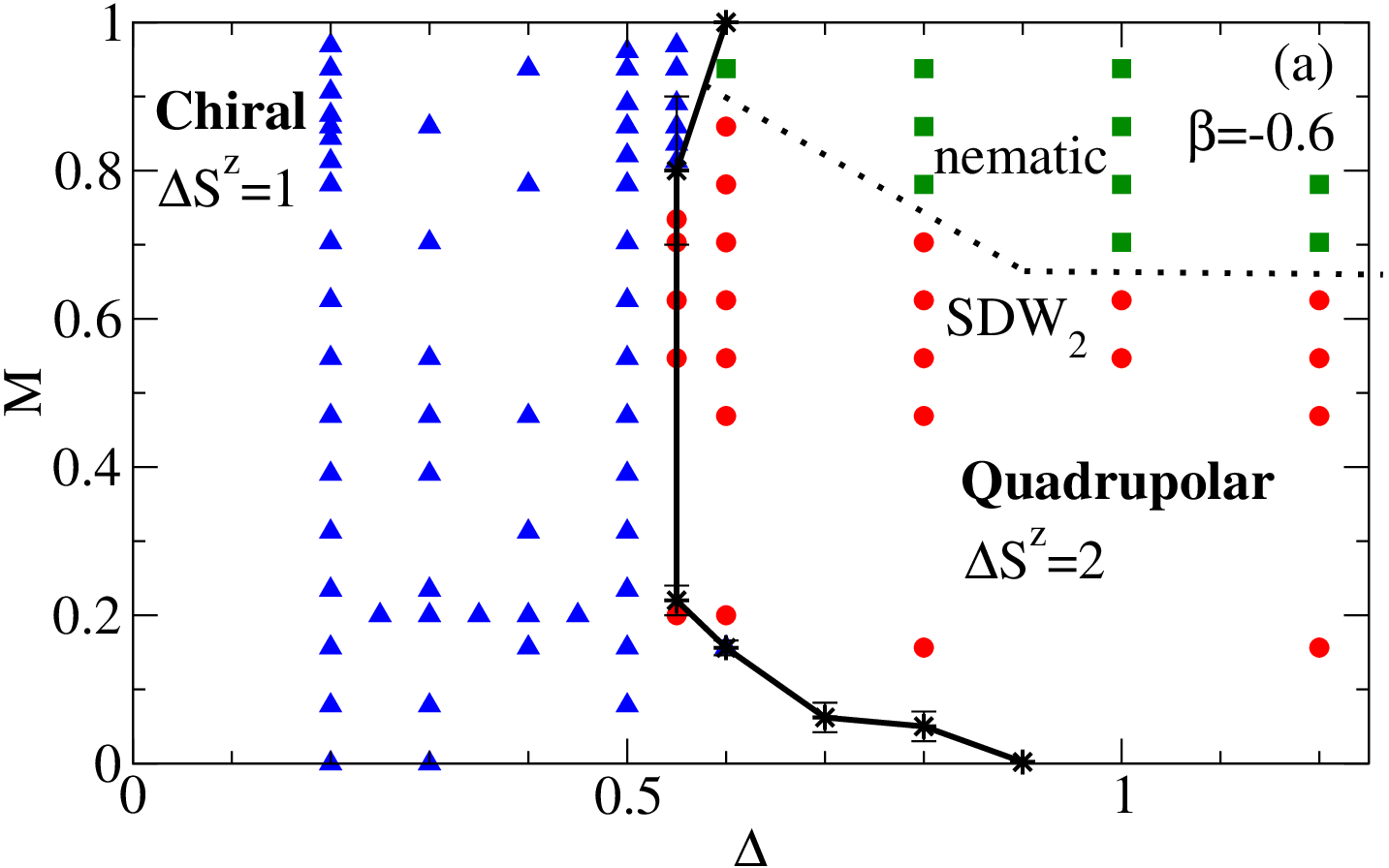}
\vspace*{2mm}
\includegraphics[width=0.45\textwidth]{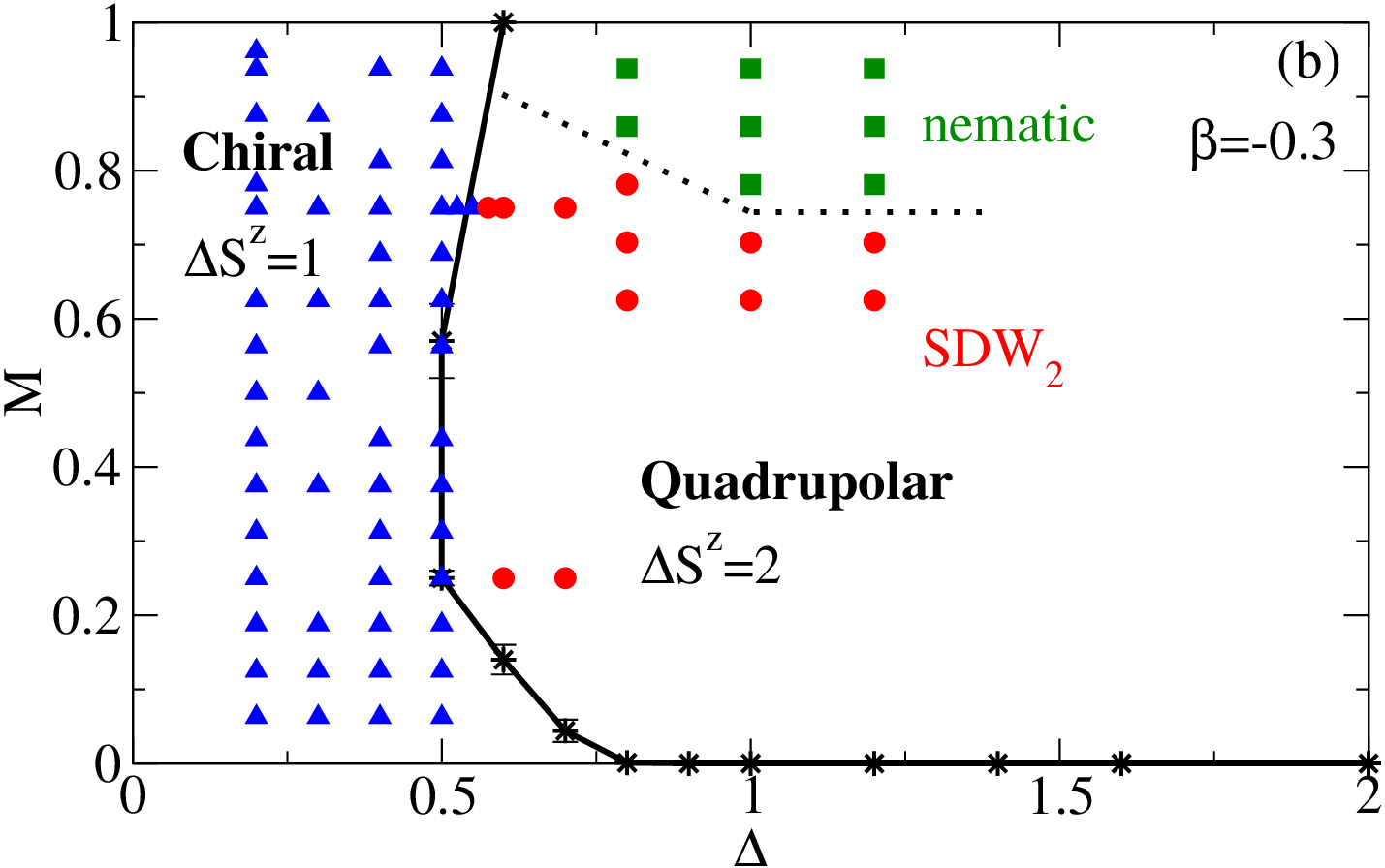}
\vspace*{2mm}
\includegraphics[width=0.45\textwidth]{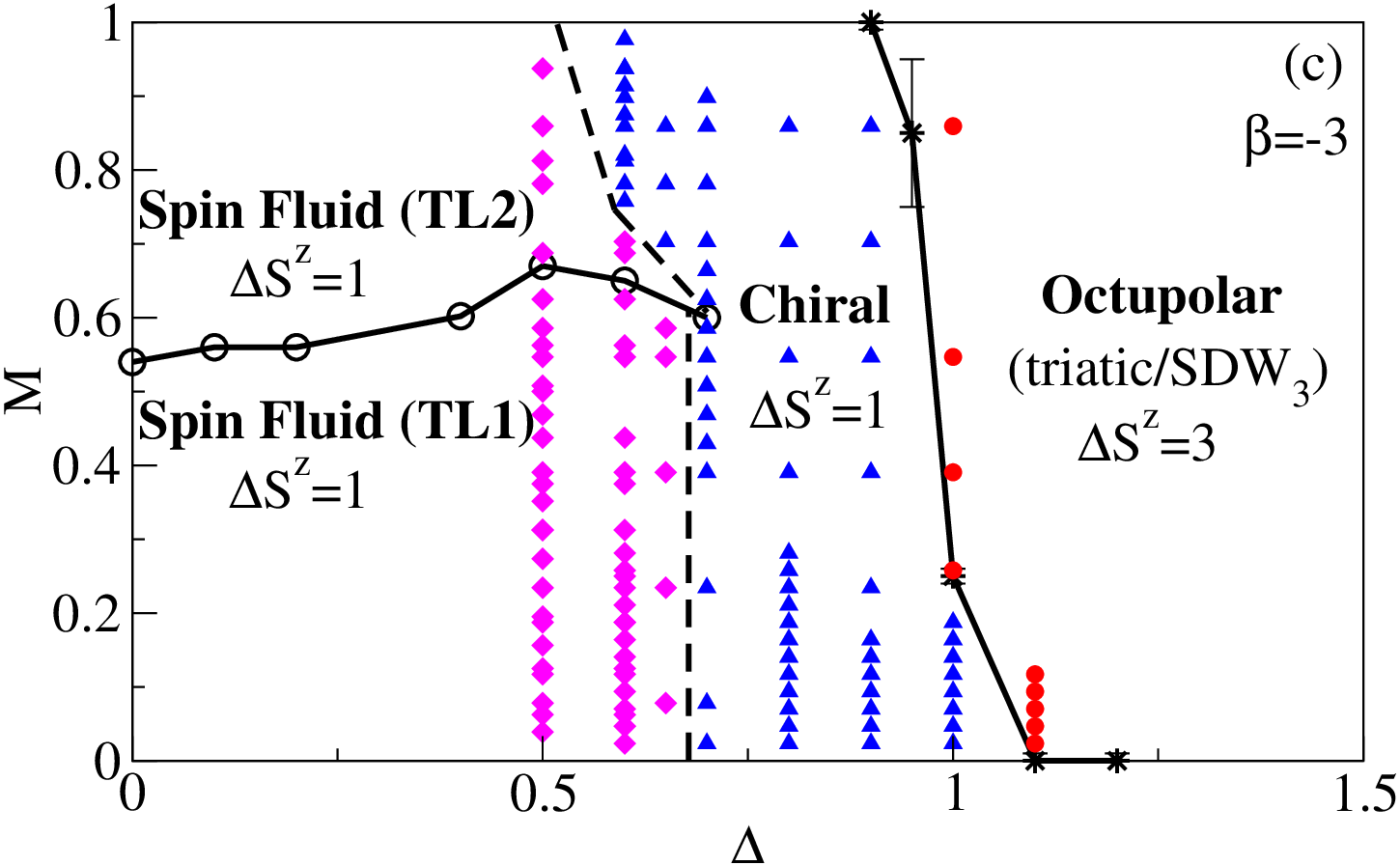}
\caption{(Color online) Phase diagrams of the Hamiltonian (\ref{ham}),
  represented as slices in the space of the magnetization $M$ and
  anisotropy $\Delta$, at fixed frustration $\beta=J_{1}/J_{2}$: (a)
  at $\beta=-0.6$; (b) $\beta=-0.3$; (c) $\beta=-3$.  Solid lines
  denote phase boundaries (extrapolated in $1/N$) as extracted from
  $M(h)$. Dashed lines denote the approximate phase boundaries as
  extracted from the correlations analysis, and dotted lines in (a)
  and (b) denote the position of a crossover between the dominant
  spin-density wave (SDW) and nematic correlations.  Solid symbols
  result from the analysis of correlations functions: Triangles (blue)
  correspond to the points in a chirally ordered phase, diamonds
  (magenta) to the one- and two-component spin-fluid (TL1 and TL2,
  respectively); at $\beta=-0.3$ and $\beta=-0.6$, circles (red)
  denote points with dominant $\rm SDW_{2}$ correlations, and squares
  (green) indicate points with dominant nematic correlations within
  the quadrupolar ($\Delta S^{z}=2$) phase.  At $\beta=-3$, circles
  (red) denote points in the octupolar ($\Delta S^{z}=3$) phase
  without specifying the dominant type of correlations ($\rm SDW_{3}$
  or triatic).  }
\label{fig:diags}
\end{figure}

\section{Magnetic phase diagrams}
\label{subsec:diags}

Summarizing all the information extracted from the analysis of magnetization
curves and correlations, one can establish the phase diagrams of the anisotropic
frustrated ferromagnetic spin-$\frac{1}{2}$ chains in the presence of a magnetic field. Such phase
diagrams in the $(M,\Delta)$ plane at different values of the frustration
$\beta=J_{1}/J_{2}$ are presented in Fig.\ \ref{fig:diags}. 

We reiterate here that we ascribe the entire $\Delta S^{z}=1$
region to the chiral phase for $\beta=-0.3$ and $\beta=-0.6$, based
 on
the very smooth character of 
how the order parameter vanishes approaching the $\Delta S^{z}=1$
boundary,  on theoretical estimates for zero-field case,\cite{furukawa08} and by
invoking 
continuity arguments. In principle,  from our data,  we cannot exclude the
existence of a small non-chiral region in the low-field part of the
phase diagram near the $\Delta S^{z}=1$
boundary, but we think that this scenario is rather unlikely. Thus, in the regime of
weakly coupled chains, the phase diagram  contains
just the chiral and quadrupolar phases, the transition between them
being likely a smooth (second-order) one.

In the strong coupling regime, $\beta=-3$, our results suggest a rich
phase diagram, displaying four phases: the octupolar, the chiral, and two
types of spin-fluid phases (which can be characterized as one- and
two-component Tomonaga-Luttinger liquids). The transition between the
chiral and the spin-fluid-TL1 phase is very sharp and is likely first order.


\section{Summary}
\label{sec:sum}

Motivated by recent experimental results for several quasi
one-dimensional magnetic
materials,\cite{hase04,enderle05,drechsler07,banks07,Buettgen+07,schrettle08,banks+09}
we  studied the model (\ref{ham}) of an anisotropic frustrated
ferromagnetic spin-$\frac{1}{2}$ chain in an external magnetic
field, at finite values of the magnetization.

We showed that an easy-axis anisotropy $\Delta>1$ stabilizes multipolar
phases,\cite{vekua07,sudan09,hikihara08} in which the total $z$-projection of
the spin $S^{z}$ increases by steps of $\Delta S^z>1$. In the presence of even
a small easy-axis anisotropy, such phases occupy the entire range of finite
magnetizations up to full saturation. Further, we found that an easy-plane
anisotropy $\Delta<1$ may favor several types of phases: chirally ordered and
non-chiral one- and two-component spin fluids.  We showed that the presence of a
moderate easy-plane anisotropy leads to the possibility of a field-induced quantum
phase transition at a substantially large value of the magnetization $M$,
even in the purely one-dimensional model (\ref{ham}), which
might provide an explanation for the field-induced 
transition\cite{banks07,schrettle08,Buettgen+07} from a helically ordered to a collinear
state observed in $\rm LiCuVO_4$.

Assuming that $\rm LiCuVO_4$ is a system of weakly coupled one-dimensional
chains and further assuming that the presence of an exchange anisotropy drives
the experimentally observed mid-field phase transition in this material, our
results imply that the low-field region would be in a helical cone-type phase
(see Ref.\ \onlinecite{UedaTotsuka09} for a recent study of helical order in a
3D magnet in high magnetic fields) while the SDW instability in the high-field
region would turn the high field region into a \emph{collinear}, magnetically
ordered state with long-range \emph{incommensurate} $\langle
S^{z}(x)S^{z}(0)\rangle$ correlations.  The former conclusion (a helical phase
in the low-field region) is in agreement with the available experimental
data,\cite{enderle05,banks07,schrettle08} while the latter conjecture of
collinear incommensurate order could be tested by neutron scattering experiments
and is consistent with the nuclear magnetic resonance measurements
\cite{Buettgen+07,schrettle08} suggesting that the magnetic order becomes
collinear in applied fields above $\approx 7.5$~T. It should be mentioned that a
similar incommensurate collinear structure has been recently observed
\cite{Kimura+08} in the quasi-1D material $\rm BaCo_{2}V_{2}O_{8}$
with easy-axis anisotropy.

For $\beta=-0.3$, our data suggest that the low-field, chiral region opens
up at a finite anisotropy; within the numerical accuracy of our calculations, we were able to resolve
the emergence of this region
 for $\Delta \lesssim 0.8$ [see Fig.~\ref{fig:diags}(b)].  

This has to be contrasted against the experimental estimate of the
easy-axis anisotropy of about 10\%,\cite{Vasilev+01,KrugVonNidda+02} and against
the fact that in the magnetization measurements in $\rm LiCuVO_4$, the mid-field
transition is observed at $7.5$~T, corresponding to about 20\% of the saturation
field.\cite{banks07,enderle05} We
stress that our results do not serve to unambiguously prove the exchange
anisotropy to be the relevant mechanism behind the mid-field transition in $\rm
LiCuVO_4$; nevertheless, our results clearly indicate that, using the values
for $\beta$ and $\Delta $ suggested for $\rm LiCuVO_4$, this material is very
close to the quantum critical point at which, as a function of decreasing
$\Delta$, a mid-field phase transition develops. This transition point shifts to
larger field as the anisotropy increases ($\Delta$ decreases).

The vicinity to many competing
phases then makes this material so interesting but also renders it difficult to
quantitatively predict its phase diagram.
 Additional experimental data
are highly desirable to clarify the nature of this phase transition,
while, in conclusion, our work shows that the emergent physics in this
model, driven by the magnetic field, quantum fluctuations and broken
exchange symmetry, is very rich.

\begin{acknowledgments}
We gratefully acknowledge fruitful discussions with S. Drechsler, M. Enderle,  A. Feiguin,
A. Honecker, A. L\"auchli, H.-J. Mikeska, L. E. Svistov, and T. Vekua. 
F.H.M.\ thanks the KITP at UCSB, where part of this
research was carried out, for its hospitality. This work was supported in part
by the National Science Foundation under Grant No.\ NSF PHY05-51164. A.K.\ was
supported by the Heisenberg Program of the Deutsche Forschungsgemeinschaft under
Grant No.\ KO~2335/1-2.
We thank E. Dagotto for granting us computing time at his group's facilities at the University of Tennessee
at Knoxville.

\end{acknowledgments}


\begin{thebibliography}{99}

\bibitem{Villain78} J. Villain, Ann. Isr. Phys. Soc. \textbf{2}, 565 (1978).


\bibitem{AndreevGrishchuk84} A. F. Andreev and I. A. Grishchuk, Sov. Phys. JETP 60, 267 (1984).


\bibitem{chubukov91a} A. V. Chubukov, Phys. Rev. B \textbf{44}, 4693 (1991).

\bibitem{kolezhuk05} A. K. Kolezhuk and T. Vekua, Phys. Rev. B \textbf{72}, 094424
  (2005).

\bibitem{laeuchli06} A. L\"auchli, F. Mila, and K. Penc,
  Phys. Rev. Lett. \textbf{97}, 087205 (2006).

\bibitem{vekua07} T. Vekua, A. Honecker, H.-J. Mikeska, and F. Heidrich-Meisner,
  Phys. Rev. B \textbf{76}, 174420 (2007).


\bibitem{hikihara08} T. Hikihara, L. Kecke, T. Momoi, and A. Furusaki,
  Phys. Rev. B \textbf{78}, 144404 (2008).


\bibitem{sudan09} J. Sudan, A. L\"uscher, and A. L\"auchli, Phys. Rev. B \textbf{80}, 140402
(2009)

\bibitem{kaburagi99} M. Kaburagi, H. Kawamura, and T. Hikihara,
  J. Phys. Soc. Jpn. \textbf{68}, 3185 (1999).

\bibitem{hikihara01} T. Hikihara, M. Kaburagi, and H. Kawamura, Phys. Rev. B
  \textbf{63}, 174430 (2001).


\bibitem{lecheminant01} P. Lecheminant, T. Jolicoeur, and P. Azaria,
  Phys. Rev. B \textbf{63}, 174426 (2001).

\bibitem{nersesyan98} A. A. Nersesyan, A. O. Gogolin, and F. H. L. E\ss{}ler,
  Phys. Rev. Lett.  \textbf{81}, 910 (1998).

\bibitem{kolezhuk00} A. K. Kolezhuk, Phys. Rev. B \textbf{62}, R6057 (2000).



\bibitem{mcculloch08} I. P. McCulloch, R. Kube, M. Kurz, A. Kleine,
  U. Schollw\"{o}ck, and A. K.  Kolezhuk, Phys. Rev. B \textbf{77}, 094404
  (2008).

\bibitem{okunishi08} K. Okunishi, J. Phys. Soc. Jpn. \textbf{77}, 114004 (2008).


\bibitem{hm06a} F. Heidrich-Meisner, A. Honecker, and T. Vekua, Phys. Rev. B
  \textbf{74}, 020403(R) (2006).


\bibitem{kecke07} L. Kecke, T. Momoi, and A. Furusaki, Phys. Rev. B \textbf{76},
  060407(R) (2007).

\bibitem{Al-Hassanieh+09} K. A. Al-Hassanieh, C. D. Batista, G. Ortiz,
  and L. N. Bulaevskii,  arXiv:0905.4871 (unpublished).

\bibitem{Bulaevskii+08} L. N. Bulaevskii, C. D. Batista, M. V. Mostovoy, and
  D. I. Khomskii,
Phys. Rev. B \textbf{78}, 024402 (2008).


\bibitem{Schulz86} H. J. Schulz, Phys. Rev. B \textbf{34}, 6372 (1986).

\bibitem{WhiteAffleck96}  S. R. White and I. Affleck, Phys. Rev. B \textbf{54}, 9862 (1996).

\bibitem{AllenSenechal97} D. Allen and D. S{\'e}n{\'e}chal, Phys. Rev. B \textbf{55}, 299 (1997).

\bibitem{ItoiQin01} C. Itoi and S. Qin, Phys. Rev. B \textbf{63}, 224423 (2001).

\bibitem{plekhanov08} E. Plekhanov, A. Avella, and F. Mancini, J. Phys.:
  Conf. Series \textbf{145}, 012063 (2009); see also
  arxiv:0811.2973 (unpublished).

\bibitem{avella08}
A. Avella, F. Mancini and E. Plekhanov,
Eur. Phys. J. B \textbf{66}, 295 (2008). 


\bibitem{somma01} R. D.  Somma and A. A. Aligia, Phys. Rev. B \textbf{64},
  024410 (2001).

\bibitem{dmitriev08} D. V. Dmitriev and V. Y. Krivnov, Phys. Rev. B \textbf{77},
  024401 (2008).

\bibitem{dmitriev09} D. V. Dmitriev and V. Y. Krivnov, Phys. Rev. B \textbf{79},
  054421 (2009).

\bibitem{furukawa08} S. Furukawa, M. Sato, Y. Saiga, and S. Onoda, J.
  Phys. Soc. Jpn. \textbf{77}, 123712 (2008).

\bibitem{enderle05} M. Enderle, C. Mukherjee, B. F{\aa}k, R. K. Kremer, J.-M. Broto,
  H. Rosner, S.-L. Drechsler, J. Richter, J. Malek, A. Prokofiev, W. A{\ss}mus,
  S. Pujol, J.-L. Raggazzoni, H. Rakoto, M. Rheinst\"adter, and H. M. R{\o}nnow,
  EPL  \textbf{70}, 237 (2005).

\bibitem{hase04} M. Hase, H. Kuroe, K. Ozawa, O. Suzuki, H. Kitazawa, G. Kido,
  and T. Sekine, Phys. Rev. B \textbf{70}, 104426 (2004).

\bibitem{drechsler07} S.-L. Drechsler, O. Volkova, A. N. Vasiliev, N. Tristan,
  J. Richter, M. Schmitt, H. Rosner, J. M\'{a}lek, R. Klingeler, A. A. Zvyagin,
  and B. B\"uchner, Phys. Rev. Lett.  \textbf{98}, 077202 (2007).


\bibitem{banks+09} M. G. Banks, R. K. Kremer, C. Hoch, A. Simon,
  B. Ouladdiaf, J.-M. Broto, H. Rakoto, C. Lee, and M.-H. Whangbo,
Phys. Rev. B  \textbf{80}, 024404 (2009).


\bibitem{banks07} M. G. Banks, F. Heidrich-Meisner, A. Honecker, H. Rakoto,
  J.-M. Broto, and R. K. Kremer, J. Phys.: Condensed Matter \textbf{19}, 145227
  (2007).

\bibitem{schrettle08} F. Schrettle, S. Krohns, P. Lunkenheimer, J. Hemberger, N. B\"uttgen, H.-A. Krug von Nidda, A. V. Prokofiev, and A. Loidl, Phys. Rev. B {\bf 77}, 144101 (2008).


\bibitem{Buettgen+07} N. B\"uttgen, H.-A. Krug von Nidda, L. E. Svistov,
  L. A. Prozorova, A. Prokofiev, and W. A{\ss}mus,
 Phys. Rev. B \textbf{76}, 014440 (2007). 



\bibitem{Vasilev+01} A. N. Vasil'ev, L. A. Ponomarenko, H. Manaka, I. Yamada,
M. Isobe, and Y. Ueda, Phys. Rev. B \textbf{64}, 024419 (2001).


\bibitem{KrugVonNidda+02} H.-A. Krug von Nidda, L. E. Svistov, M. V. Eremin,
  R. M. Eremina, A. Loidl, V. Kataev, A. Validov, A. Prokofiev, and
  W. A{\ss}mus, Phys. Rev. B \textbf{65}, 134445 (2002).


\bibitem{white92b} S. R. White, Phys. Rev. Lett. \textbf{69}, 2863 (1992).

\bibitem{white93} S. R. White, Phys. Rev. B \textbf{48}, 10345  (1993).

\bibitem{schollwoeck05} U. Schollw\"ock, Rev. Mod. Phys. \textbf{77}, 259
  (2005).



\bibitem{okunishi03} K. Okunishi and T. Tonegawa,
  J. Phys. Soc. Jpn. \textbf{72}, 479 (2003).

\bibitem{note1} For the zero-field case, the results of Ref.~\protect\onlinecite{somma01}
imply the antiferromagnetic (``up-up-down-down'' type) order for $\beta
\lesssim -3$ and $\Delta>1$. We have not attempted to make a connection
to these results as our key interest is in the finite magnetization case, where
the antiferromagnetic order is already destroyed.

\bibitem{KuzianDrechsler07} R. O. Kuzian and S.-L. Drechsler,
  Phys. Rev. B \textbf{75}, 024401 (2007).

\bibitem{okunishi99} K. Okunishi, Y. Hieida, and Y. Akutsu, Phys. Rev. B
  \textbf{60}, R6953 (1999).

\bibitem{McCulloch07} I. P. McCulloch, J. Stat. Mech.: Theor. Exp., P10014 (2007).

\bibitem{iDMRG} I. P. McCulloch,  arXiv:0804.2509 (unpublished).

\bibitem{criticalscaling}T. Nishino, K. Okunishi and M. Kikuchi, Phys. Lett. A
  \textbf{213}, 69 (1996); L. Tagliacozzo, Thiago R. de Oliveira, S. Iblisdir,
  and J. I. Latorre, Phys. Rev. B \textbf{78}, 024410 (2008).

\bibitem{Pollmann}F. Pollmann, S. Mukerjee, A. Turner, J. E. Moore,
  Phys. Rev. Lett. \textbf{102}, 255701 (2009).

\bibitem{IanInProgress}I. P. McCulloch, L. Tagliacozzo and G. Vidal, in preparation.

\bibitem{note-errbars} The error bars in the finite-size DMRG results
  for the chiral order parameter
  $\kappa_{0}^{2}$ in Figs.\ \ref{fig:chirality1},\ref{fig:chirality3} are
  crude estimates reflecting only the degree of asymptotic ``flatness'' of the
  chiral correlators like shown in Fig.\ \ref{fig:kz-corr}  (larger
  error bars imply that there is still a nonzero slope at large
  distances). Those error bars are thus just the measure of the
  strength of  proliferation of the boundary effects into the bulk, 
and do not include any uncertainties related
  to the finite-size effects.






\bibitem{note4}
The reason is that a too small $\xi$ effectively introduces a gap in the
symmetric sector, which kills the twist term\protect\cite{nersesyan98} and prevents the
emergence of chirality.

\bibitem{UedaTotsuka09} H. T. Ueda and K. Totsuka, Phys. Rev. B \textbf{80}, 014417 (2009).

\bibitem{Kimura+08} S. Kimura, M. Matsuda, T. Masuda, S. Hondo, K. Kaneko,
  N. Metoki, M. Hagiwara, T. Takeuchi, K. Okunishi, Z. He, K. Kindo,
  T. Taniyama, and M. Itoh, Phys. Rev. Lett. \textbf{101}, 207201 (2008).








\end{thebibliography}

\end{document}